\documentclass{aa}  
\bibpunct{(}{)}{;}{a}{}{,} 

\usepackage{graphicx}
\usepackage{txfonts}
\def\kms{km$\,$s\ensuremath{^{-1}}}
\def\aFe{[$\alpha/{\rm Fe}$]~}
\def\Ha{${\rm H}{\alpha}$}
\def\Hb{${\rm H}{\beta}$}

\def\hi{H~{\small I}}
\def\Mgb{{\rm Mg}$b$}
\def\Mg2{{\rm Mg}$_2$}
\def\Fe{$\langle {\rm Fe}\rangle$}

\def\ZH{[$Z/{\rm H}$]~}
\def\MgFe{[${\rm MgFe}$]$'$}

\def\oiiipg{[O~{\small III}]$\,\lambda\lambda4959,5007$}

\def\oiiig{[O~{\small III}]$\,\lambda5007$}

\def\niipg{[N~{\small II}]$\,\lambda\lambda6548,6583$}

\def\niig{[N~{\small II}]$\,\lambda6583$}
\def\nii{[N~{\small II}]}
\def\siipg{[S~{\small II}]$\,\lambda\lambda6716,6731$}

\def\sii{[S~{\small II}]}

\begin{document}

\title{The external origin of the polar gaseous disk of the S0 galaxy IC 5181
  \thanks{Based on observation collected at the European Southern
    Observatory for the programme 63.N-0327(A).}  
 \thanks{Tables 1---3 are available in electronic form at http://www.aanda.org.}}


\author{A. Pizzella \inst{1,2} 
        \and 
        L. Morelli\inst{1,2}
        \and
        E. M. Corsini\inst{1,2}
        \and
        E. Dalla Bont\`a\inst{1,2}
        \and
        M. Cesetti\inst{1}
        }

\institute{Dipartimento di Fisica e Astronomia ``G. Galilei'',
  Universit\`a di Padova, vicolo dell'Osservatorio 3, I-35122 Padova,
  Italy\\ 
\email{alessandro.pizzella@unipd.it} 
\and INAF-Osservatorio Astronomico di Padova, vicolo dell'Osservatorio
5, I-35122 Padova, Italy.}

\date{Received July 5, 2013; accepted September 19, 2013}

\abstract
%
{Galaxies accrete material from the environment through acquisition
  and merging events. These processes contribute to galaxy assembly
  and leave their fingerprints on the galactic morphology, internal
  kinematics of gas and stars, and stellar populations.}
%
%
{We study the nearby S0 galaxy IC~5181 to address the origin of the
  ionized-gas component orbiting the galaxy on polar orbits.}
%
%
{We measure the surface brightness distribution of the stars and
  ionized gas of \object{IC~5181} from broad and narrow-band imaging. The
  structural parameters of the galaxy are obtained with a photometric
  decomposition assuming a S\'ersic and exponential profile for the
  bulge and disk, respectively. We measure the ionized-gas and stellar
  kinematics and the line strengths of the Lick indices of the stellar
  component along both the major and minor axis. The age, metallicity,
  and \aFe\ enhancement of the stellar populations are derived using
  single stellar population models with variable element abundance
  ratios. The ionized-gas metallicity is obtained from the equivalent
  width of the emission lines.}
%
%
{IC~5181 is a morphologically undisturbed S0 galaxy with a classical
  bulge made by old stars with super solar metallicity and
  overabundance. Stellar age and metallicity decrease in the disk
  region. The galaxy hosts a geometrically and kinematically decoupled
  component of ionized gas. It is elongated along the galaxy minor
  axis and in orthogonal rotation with respect to the galaxy disk.}
%
%
{ We interpret the kinematical decoupling as suggestive of a component 
of gas, which is not related to the stars and having an external
  origin. It was accreted by IC~5181 on polar orbits from the
  surrounding environment.}


\titlerunning{Gas on polar orbits in IC 5181}

\authorrunning{A. Pizzella et al.}

\maketitle

%

\section{Introduction}
\label{sec:intro}

It is widely accepted that galaxies undergo a number of acquisition
and merging events during their assembly and growth. Depending on the
nature and mass of the accreted material and the geometry and duration
of the accretion process, such events may leave temporary or permanent
signatures in the morphology, kinematics, and stellar populations of
the host galaxy (see the reviews by \citealt{1994AJ....108..456R},
\citealt{1998giis.conf..105S}, and \citealt{1999IAUS..186..149B}).
Some examples of the features observed in lenticular and spiral
galaxies are the counter-rotating disks of gas
\citep[][]{1995Natur.375..661C, 2012MNRAS.422.1083C} and stars
\citep{1992ApJ...394L...9R, 1996ApJ...458L..67B}, nuclear polar disks
\citep{2003A&A...408..873C, 2004AJ....127.2641S} and large-scale polar
rings \citep{1990AJ....100.1489W, 2011MNRAS.418..244M}, polar bulges
\citep{1999ApJ...519L.127B, 2004AJ....128..137M}, 
and nuclear stellar disks
\citep{2004A&A...428..877K, 2012MNRAS.423L..79C}.

Lenticular galaxies are gas-poor systems \citep{1984ARA&A..22..445H}.
Several pieces of evidence suggest that the origin of gas is
external, in at least a fair fraction of S0s, and not only in some peculiar cases. 
Assuming gas infall with randomly oriented angular momentum,
\citet{1992ApJ...401L..79B} found that the ionized gas is of external
origin in about $40\%$ of the S0 galaxies in their sample. This result
has been confirmed by \citet{1996MNRAS.283..543K} who set the
fraction of S0s with retrograde gas to $24\pm10\%$. More recently,
studying a large sample of fast-rotating early-type galaxies,
\citet{2011MNRAS.417..882D} found that $42\pm5\%$ of galaxies show
decoupling between the stellar and gaseous components.

Counter-rotating gaseous disks are observed in less than $12\%$ of
spiral galaxies and only a few of them host a significant fraction of
counter-rotating stars \citep{2001AJ....121..140K,
  2004A&A...424..447P}.  The comparison with S0 galaxies suggests that
the retrograde acquisition of small amounts of external gas gives rise
to counter-rotating gaseous disks only in S0s, while in gas-rich
spirals the newly acquired gas is swept away by the pre-existing gas.
The gas from the environment interacts and dissipates with that of the
host galaxy and in most of cases it will not survive to
build a counter-rotating component. Indeed, counter-rotating gaseous
and stellar disks in spirals are formed only from the retrograde
acquisition of large amounts of gas exceeding that of pre-existing gas
and fuelling a subsequent in situ star formation.  The two
counter-rotating stellar components are expected to have different
stellar populations \citep{2013A&A...549A...3C, 2013ApJ...769..105K}.

In addition, there are disk galaxies where the angular momenta of the
main stellar body and decoupled gaseous component are even orthogonal
to each other. This is the case of polar ring galaxies. The accretion
of external material through the capture of gas clouds or merging
\citep{2010ApJ...714.1081S, 2013arXiv1302.7273C} is invoked also to
explain the formation of this class of relatively rare objects
\citep{2011MNRAS.418..244M}. The infall of massive gas clouds gives
rise to the formation of large-scale polar disks of stars
\citep{2002A&A...391..103I}. Inclined disks of stars and gas are also
observed in lenticular galaxies \citep{2009ApJ...694.1550S}.
The stability of polar and inclined orbits in axisymmetric and
triaxial potentials have been addressed both from the theoretical
\citep[e.g.,][]{1993A&A...268...65F, 2003MNRAS.341.1179A} and
observational point of view 
\citep{1997A&A...323..349P, 2006MNRAS.365..367E, 2006MNRAS.370..753B}. 

\object{IC~5181} is a large \citep[$2\farcm6\, \times\, 0\farcm8$;][hereafter
  RC3]{1991rc3..book.....D}, and bright ($B_T=12.51$; RC3) early-type
disk galaxy. It is classified as edge-on SA0 in RC3 and S0$_1$(7) in
\citet{1994cag..book.....S} due to the presence of a flattened thick
disk. Its total absolute magnitude is $M^0_{B, {\rm T}} = -19.48$
corrected for inclination and extinction (RC3) and adopting a
distance of 24.8 Mpc \citep{1988ngc..book.....T}.  IC~5181 is a member
of the loose NGC~7213 group \citep{1989ApJS...69..809M,
  1993A&AS..100...47G}. It forms a pair with the edge-on spiral galaxy
NGC~7232A at $8\farcm1$ separation corresponding to a projected linear
distance of 58.4 kpc.

In the framework of acquisition events occurred in the lifetime of
galaxies, we present IC~5181 as a new case of a disk galaxy
characterized by a geometric and kinematic orthogonal decoupling
between its stellar body and the ionized-gas component. This paper is
organized as follows. The photometric and spectroscopic observations
of IC~5181 are presented in Sect. \ref{sec:data}. The results about
the measurements of the distribution, kinematics, and chemical
properties of the stars and ionized gas are discussed in
Sect. \ref{sec:results}. The conclusions are given in
Sect. \ref{sec:conclusions}.

%

\section{Observations and data reduction}
\label{sec:data}

%

\subsection{Narrow and broad-band imaging}
\label{sec:imaging}

The photometric observations of IC~5181 were carried out at European
Southern Observatory (ESO) in La Silla with the 2.2-m MPG/ESO
telescope equipped with the Wide Field Imager (WFI) on July 22, 1999.

The galaxy was centered on the No. 51 EEV CCD which has $2048\,
\times\, 4096$ pixels of $15\, \times\, 15$ $\rm \mu m^2$. It yielded
a field of view of $8\farcm1\, \times\, 16\farcm2$ with an image scale
of $0\farcs238$ pixel$^{-1}$. The gain and readout noise are 2.1 $e^-$
count$^{-1}$ and 5.2 $e^-$ (rms), respectively.  Two 300-s narrow-band
images were obtained with the \Ha$/7$ No. 856 filter in order to
isolate the spectral region characterized by the redshifted \Ha\ and
\niipg\ emission lines according to the galaxy systemic velocity. One
30-s, two 300-s, and one 600-s images were taken with the Ic/Iwp
No. 845 broad-band filter in order to subtract off the stellar
continuum from the emission-band images.  The shortest exposure was
taken to deal with the saturation of the galaxy nucleus resulting in
the deeper broad-band exposures. Since the night was characterized by
good photometric conditions, some photometric standards were
observed to calibrate the flux of broad-band images.

The data reduction was performed using standard MIDAS\footnote{The
  Munich Image Data Analysis System (MIDAS) is developed and
  maintained by the European Southern Observatory (ESO).} routines.
All the images were bias subtracted and flat-field corrected. The sky
level was determined in a number of regions selected to be empty areas
far from IC~5181 to avoid the contamination of the target galaxy as
well as of foreground and background sources. The broad-band images
were affected by fringing, which was corrected by subtracting the
fringing-pattern frame made available to this aim by the ESO
calibration plan of WFI. This reduced the fringing level from $5\%$ to
less than $1\%$ of the sky level. No fringing correction was needed
for the narrow-band images. Then the images were shifted and aligned
to an accuracy of a few hundredths of a pixel using common field stars
as a reference.  After checking their point-spread functions (PSFs)
were comparable, the frames obtained with the same filter were combined
to obtain a single narrow and single broad-band image. The cosmic rays
were identified and removed during the averaging process. The seeing
{\em FWHM\/} of the resulting narrow and broad-band image as measured
by fitting a two-dimensional Gaussian to the field stars is $0\farcs7$
and $0\farcs6$, respectively. The broad-band image was flux calibrated
to the Cousins $I$-band.  The photometric calibration constant
includes only the correction for atmospheric extinction, which is
taken from the differential aerosol extinction for ESO
\citep{1995Msngr..80...34B}.  No colour term has been considered and no
attempt was made to correct for internal and Galactic extinction.

Finally, the broad-band image was convolved with a Gaussian PSF to
yield the same PSF {\em FWHM\/} of the narrow-band image. The
broad-band image was suitably scaled and subtracted from the
narrow-band image to obtain a continuum-free map of the
\Ha$+$\nii\ emission of the galaxy. The scale factor was estimated by
comparing the surface brightness measured in the two pass-bands
between $20\arcsec$ and $30\arcsec$ from the center.

\begin{figure}
 \centering
 \includegraphics[width=9cm]{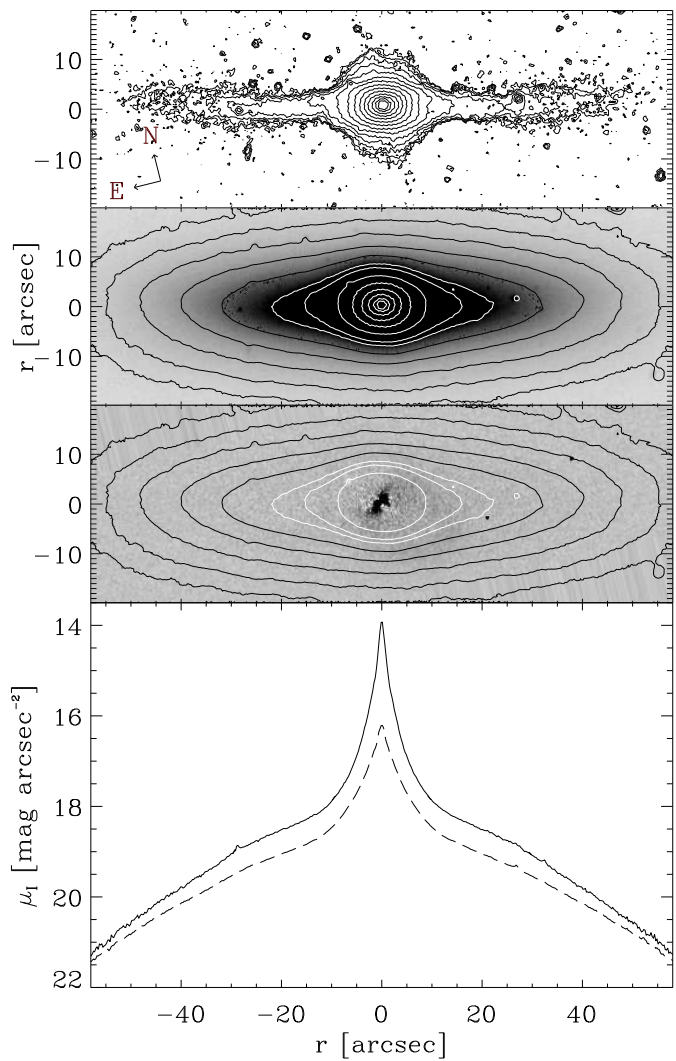}
 \caption{Surface-brightness distribution of IC~5181. The
   unsharp-masked image, the continuum-band image with some $I-$band
   isophotal contours (ranging from 14.7 to 21.6 mag arcsec$^{-2}$
   with increment of 0.6 mag arcsec$^{-2}$), the continuum-free \Ha$+$\nii\ image
   with $I-$band isophotal contours (same levels as above), and the
   radial profiles of the surface brightness extracted along the major
   axis (solid line) and over a rectangular aperture with a width of
   $40\arcsec$ parallel to the major axis and centered on the galactic
   nucleus (dashed line) are shown (from top to bottom). The
   orientation of the field of view is given in the top panel. }
 \label{fig:unsharp}

\end{figure}

The resulting $I$-band and continuum-free images of IC~5181 are shown
in Fig.~\ref{fig:unsharp}.

%

\subsection{Long-slit spectroscopy}
\label{sec:spectroscopy}

The spectroscopic observations of IC~5181 were carried out at ESO in
La Silla with the 1.52-m ESO telescope on June 9-11, 1999.
The telescope was equipped with the Cassegrain Boller \& Chivens
spectrograph. The No. 33 grating with 1200 grooves mm$^{-1}$ was used
in the first order in combination with a $2\farcs2\, \times\,
4\farcm2$ slit, and the No. 39 Loral/Lesser CCD which has $2048\,
\times\, 2048$ pixels of $15\, \times\, 15$ $\rm \mu m^2$. The gain
and readout noise are 1.2 $e^-$ count$^{-1}$ and 5.4 $e^-$ (rms),
respectively.
The spectral range between about $4850$ \AA\ and $6850$ \AA\ was
covered with a reciprocal dispersion of $0.98$ \AA\ pixel$^{-1}$. The
spatial scale was $0\farcs81$ pixel$^{-1}$.
The instrumental resolution was 2.75 \AA\ ({\em FWHM\/}) and it was
derived as the mean of the Gaussian {\em FWHM\/}'s measured for a
dozen unblended arc-lamp lines distributed over the whole spectral
range of a wavelength-calibrated comparison spectrum.  It corresponds
to $\sigma_{\rm instr} = 53$ \kms\ at \Ha.

\begin{figure}
 \centering
 \includegraphics[width=9cm]{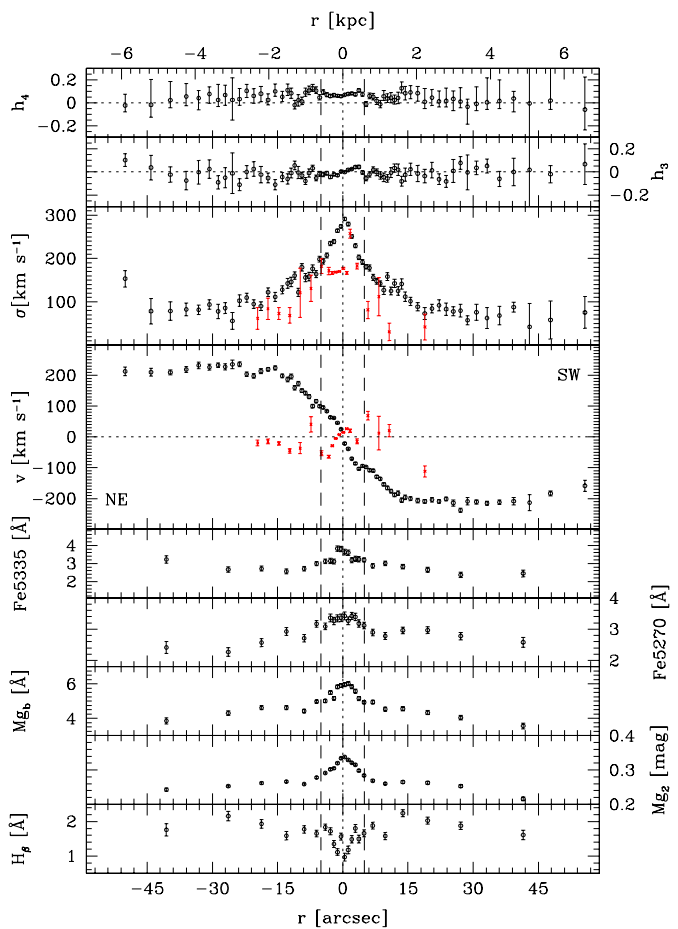}
 \caption{Kinematic parameters of the stars (circles) and ionized gas
   (crosses) and the line-strength indices measured along the major
   axis of IC~5181 ({\rm PA = 74\degr}). The radial profiles of the
   line-of-sight fourth- and third-order coefficient of the
   Gauss-Hermite decomposition of the LOSVD ($h_4$ and $h_3$),
   velocity ($v$) after the subtraction of systemic velocity, velocity
   dispersion ($\sigma$), and of the line-strength indices Fe5335,
   Fe5270, \Mgb , \Mg2, and \Hb\ are plotted (from top to
   bottom). The vertical dashed lines correspond to the radii ($|r| =
   r_{\rm bd}$), where the surface-brightness contributions of the
   bulge and disk are equal.}
 \label{fig:major}
\end{figure}
\begin{figure}
 \centering
 \includegraphics[width=9cm]{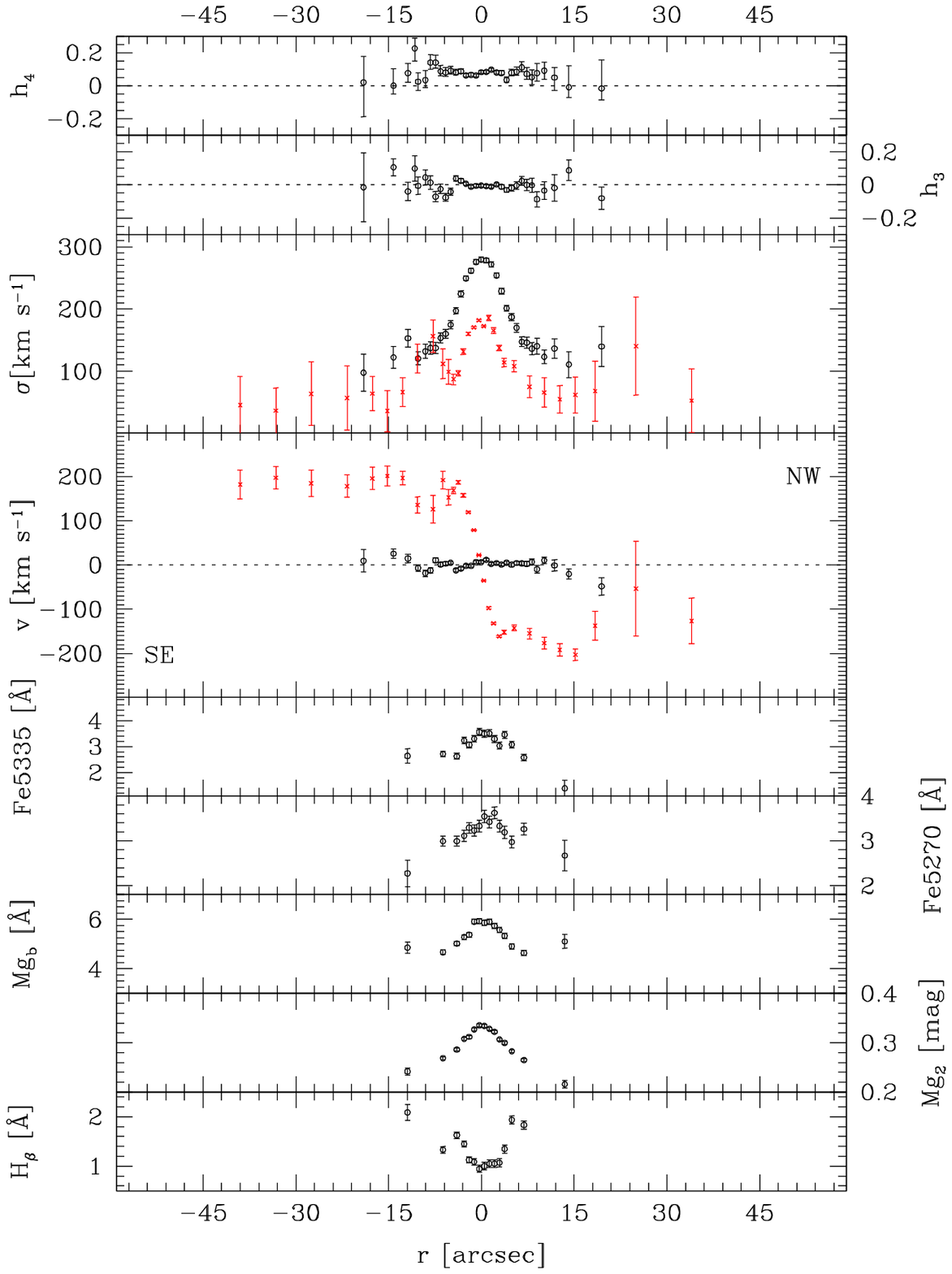}
 \caption{As in Fig. \ref{fig:major}, but for the minor axis of IC~5181
   ({\rm PA = 164\degr}).}
 \label{fig:minor}
\end{figure}

IC~5181 was observed along the major ($\rm PA = 74\degr$) and minor
axis ($\rm PA = 164\degr$). At the beginning of each exposure the
galaxy was centered on the slit using the guiding camera. Repeated
exposures of 2700 s each ensured 1.5 and 3.0 hours of effective
integration along the major and minor axis, respectively without
storing up too many cosmic rays. Some spectro-photometric standard
stars were observed to calibrate the flux of the spectra before
line-strength indices were measured. Spectra of the comparison arc
lamp were taken before and/or after object exposures.
The value of the seeing {\em FWHM} during the observing nights,
measured by the La Silla Differential Image Motion Monitor (DIMM)
ranged between $1\farcs0$ and $1\farcs5$.

All the spectra were bias subtracted, flat-field corrected, cleaned of
cosmic rays, corrected for bad columns, and wavelength and flux
calibrated using MIDAS. Each spectrum was rebinned using the
wavelength solution obtained from the corresponding arc-lamp spectrum.
All the galaxy and stellar spectra were corrected for CCD
misalignment. The sky contribution was determined by interpolating
along the outermost $10\arcsec - 30\arcsec$ at the two edges of the slit, where
the galaxy or stellar light was negligible, and then subtracted. A sky
subtraction better than $1\%$ was achieved. Each spectrum was
flux-calibrated using the sensitivity function obtained from the flux
standard star spectrum of the corresponding night. The spectra
obtained for the same galaxy along the same axis were coadded using
the center of the stellar continuum as a reference, thus improving the
signal-to-noise ratio ($S/N$) of the final two-dimensional spectrum.

The stellar kinematics was measured from the galaxy absorption
features present in the wavelength range and centered on the
Mg~{\small{I}} line triplet ($\lambda\lambda\,$ 5164, 5173, 5184 \AA)
using the Gas and Absorption Line Fitting \citep[GANDALF;
][]{2006MNRAS.366.1151S} IDL\footnote{The Interactive Data Language
  (IDL) is distributed by ITT Visual Information Solutions.} code
adapted for dealing with the spectra of the 1.52-m ESO telescope.
The galaxy spectra were rebinned along the dispersion direction to a
logarithmic scale, and along the spatial direction to obtain a $S/N
\geq 25$ per resolution element.
At each radius a linear combination of template stellar spectra from
the MILES library by \citet{2006MNRAS.371..703S} was convolved with
the line-of-sight velocity distribution (LOSVD) and fitted to the
observed galaxy spectrum by $\chi^2$ minimization in pixel space.
The LOSVD was assumed to be a Gaussian plus third- and fourth-order
Gauss-Hermite polynomials, which describe the asymmetric and symmetric
deviations of the LOSVD from a pure Gaussian profile 
\citep{1993ApJ...407..525V, 1993MNRAS.265..213G}.
This allowed us to derive the radial profiles of the line-of-sight
velocity, velocity dispersion, and third- and fourth-order
Gauss-Hermite moments of the stellar component.

The ionized-gas kinematics was measured with GANDALF from the emission
lines present in the spectra, namely \Hb, \oiiipg, \niipg, \Ha, and
\siipg . The low $S/N$ of the stellar continuum prevented us to use
GANDALF for measuring the gas kinematics for $|r>17\arcsec|$ along the
galaxy minor axis. In this radial range the lines of the \nii\ doublet
and \Ha\ were fitted by Gaussians, while describing the stellar
continuum with a low-order polynomial. We averaged adjacent spectral
rows to increase the $S/N$ of the relevant emission lines.  The
Gaussians were assumed to share the same velocity and velocity
dispersion which correspond to the line-of-sight velocity and velocity
dispersion of the ionized gas, respectively. A flux ratio of 1:2.96
was assumed for the \nii\ doublet, as dictated by atomic physics 
\citep[e.g.,][]{1989agna.book.....O}. The
best-fitting Gaussian parameters were derived using a non-linear
least-squares minimization based on the robust Levenberg-Marquardt
method using the MPFIT algorithm \citep{2009ASPC..411..251M}
under the IDL environment as done in
\citet{2008MNRAS.387.1099P}. This allowed us to trace the radial
profiles of the line-of-sight velocity and velocity dispersion of the
ionized-gas component.

The uncertainties on the kinematic parameters of the stars and ionized
gas were estimated by rescaling the formal errors given as outputs
from the least-squares fitting routines to a reduced $\chi^2=1$.

The Mg, Fe and \Hb\ line-strength indices along the major and minor
axes were measured following \citet{2012MNRAS.423..962M}. The
flux-calibrated spectra were rebinned in the radial direction to
achieve a $S/N \geq 50$ per resolution element. The spatial binning is
therefore different from the one used for the kinematical
measurements.
The average iron index $\rm{\left<Fe\right> = (Fe5270 + Fe5335)/2}$
\citep{1990MNRAS.245..217G}, and the combined magnesium-iron index
$[{\rm MgFe}]^{\prime}=\sqrt{{\rm Mg}\,b\,(0.72\times {\rm Fe5270} +
  0.28\times{\rm Fe5335})}$ \citep{2003MNRAS.339..897T} were computed
too. Errors on indices were derived from photon statistics and CCD
readout noise, and calibrated by means of Monte Carlo simulations.
The contamination of the \Hb\/ line-strength index by the \Hb\/
emission line due to the ionized gas present in the galaxy is a
problem when deriving the properties of the stellar populations. To
address this issue, the \Hb\/ index was measured from the galaxy
spectrum after subtracting the contribution of the \Hb\ emission
line. Only \Hb\/ emission lines detected with a $S/N > 3$ were
subtracted from the observed spectra.

The stellar and ionized-gas kinematics as well as the line-strength
indices measured along the major and minor axis of IC~5181 are plotted
in Fig.~\ref{fig:major} and Fig.~\ref{fig:minor}, respectively.  The
kinematical measurements are available in Tables 1 and 2 for the
stellar and ionized gas, respectively. The line-strength indices
measurements are available in Table 3.

%

\section{Results}
\label{sec:results}

%

\subsection{Surface photometry}
\label{sec:photometry}

\begin{figure}
  \centering
  \includegraphics[width=9cm]{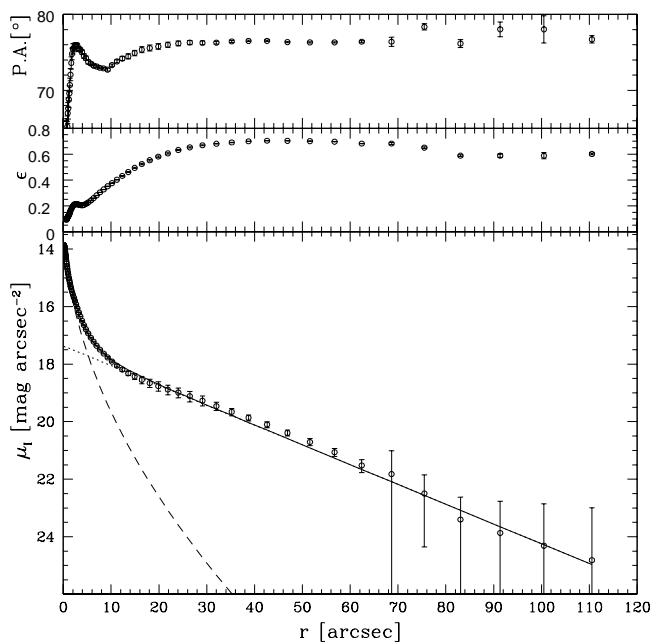}
  \caption{Isophotal parameters as a function of the semi-major-axis
    distance and photometric decomposition of IC~5181. The radial
    profiles of position angle (PA), ellipticity ($\epsilon$), and
    Cousins $I$-band surface brightness ($\mu_{I}$) are plotted (from
    top to bottom). The dashed, dotted and solid lines represent the
    surface-brightness radial profiles of the bulge, disc, and model
    obtained from the photometric decomposition, respectively.}
  \label{fig:ellipse}
\end{figure}

Isophote-fitting with ellipses, after masking foreground stars and
residual bad columns, was carried out on the broad-band image using
the IRAF\footnote{The Imaging Reduction and Analysis Facility (IRAF)
  is distributed by the National Optical Astronomy Observatory, which
  is operated by the Association of Universities for Research in
  Astronomy (AURA), Inc., under cooperative agreement with the
  National Science Foundation.}  task ELLIPSE.  First isophotes were
fitted by ellipses allowing their centers to vary. Within the errors,
no variation in the ellipses center was found. Therefore, the final
ellipse fits were done at a fixed ellipse center out to a distance of
$110\arcsec$ from the center where the galaxy surface brightness is larger
than 0.5 times the standard deviation of the sky surface brightness.
The ellipse-averaged profiles of surface brightness, position angle,
and ellipticity are plotted in Fig.~\ref{fig:ellipse}.

The position angle of the
isophotes peaks to $76\degr$ at $3\arcsec$ from the center. At this
radius the ellipticity peaks at $0.22$. Further out, the position
angle and ellipticity decrease to $\rm PA = 73\degr$ at $10\arcsec$
and $\epsilon = 0.20$ at $4\arcsec$, respectively.  At larger radii,
both the position angle and ellipticity smoothly increase. The
position angle reaches a constant value $76\degr$ at radii larger than
$25\arcsec$, the ellipticity rises to a maximum $0.70$ at $45\arcsec$
and then decreases to $0.60$ at the outermost observed point.
The total magnitude of IC~5181 ($I_T = 10.07\pm0.18$) was obtained by
extrapolating the surface-brightness radial profile. Our measurements
are fully consistent with those by \citet{2011ApJS..197...21H}.

The observed change in the position angle of the isophotes between
$3\arcsec$ and $15\arcsec$ is not expected for an S0 galaxy hosting an
axisymmetric bulge. Therefore, a two-dimensional photometric
decomposition of the surface-brightness distribution of IC~5181 was
performed to better investigate the structure of the galaxy. Two
different parametric fitting algorithms were used to this aim, but
neither GALFIT \citep{2010AJ....139.2097P} nor GASP2D
\citep{2008A&A...478..353M} returned a reliable description of the
galaxy. Adding one or more extra axisymmetric components or adopting a
thick disk do not improve the result. This suggests the presence of a
non-axisymmetric component (either a bar or a triaxial bulge) which
can not be properly decomposed due to the high inclination of the
galaxy.

Following \citet{2006MNRAS.370..753B}, the unsharp-masked image of the
broad-band frame was built to first gauge the structure and extent of
such a non-axisymmetric structure (Fig.~\ref{fig:unsharp}).  The
unsharp-mask image was obtained by median-filtering the broad-band
image, that is, by replacing the value of each pixel by the difference
between it and that of the median within a centered circular aperture
of $9\arcsec$. The size of the aperture was held fixed across the
image, but it was chosen to best highlight the features of
interest. The surface-brightness distribution of the unsharp-masked
image is remarkably similar to that of the median-filtered image of
the $N$-body simulations performed by \citet{2005MNRAS.358.1477A} to
investigate to light distribution of a strong bar seen end on.

In addition, the continuum-subtracted image shows the presence of a
faint and featureless ionized-gas emission within the inner $6\arcsec$
from the galaxy center (Fig.~\ref{fig:unsharp}).  The flux of the
ionized-gas emission is only a few percents with respect to that of
the underlying continuum and the gaseous structure is elongated almost
perpendicularly with respect to the galaxy major axis.

We decided to perform a standard one-dimensional photometric
decomposition similar to that adopted by several authors
\citep[e.g.,][]{1977ApJ...217..406K, 2001A&A...367..405P,
  2008A&A...478..353M} in order to interpret the radial trends of the
kinematic and chemical properties measured for the stars and ionized
gas in IC~5181. The radial profile of the surface brightness was
modeled as the sum of the contribution of a S\'ersic bulge and an
exponential disk.
The effective surface brightness $\mu_{\rm e}$, effective radius
$r_{\rm e}$, and shape parameter $n$ of the bulge and the central
surface brightness $\mu_0$ and scale-length $h$ of the disk were
estimated by $\chi^2$ minimization of the surface brightnesses in
counts pixel$^{-1}$ using the MPFIT algorithm. Each data point was
weighted according to the variance of its total observed counts due to
the contribution of both galaxy and sky, and determined assuming
photon noise limitation and taking the detector gain and read-out
noise into account. Seeing effects were also taken into account by
convolving the model surface brightness with a Gaussian PSF with an
{\em FWHM\/} matching the observed one. The convolution was performed
as a product in Fourier domain before the least-squares
minimization. The uncertainties on the structural parameters were
estimated by rescaling the formal errors of the fit to a reduced
$\chi^2=1$.
The best-fitting values and their $1\sigma$ errors are $\mu_{\rm e} =
16.8 \pm 0.1$ mag~arcsec$^{-2}$, $r_{\rm e} = 3\farcs4 \pm 1\farcs4$,
$n = 1.89 \pm 0.09$ for the bulge and $\mu_0 = 17.4 \pm 0.1$
mag~arcsec$^{-2}$, $h = 15\farcs8 \pm 0\farcs2$ for the disk.
The bulge-to-disk luminosity ratio $B/D = 0.6$ was derived from the
total luminosity after subtracting the disk contribution. 
The bulge and disk provide the same contribution to the
total surface brightness at $r_{\rm bd} = 5\arcsec$.
The result of the photometric decomposition is shown in
Fig.~\ref{fig:ellipse}.
As far as the other structural parameters of the galaxy concern, the
innermost peak values of the isophotal position angle and ellipticity
were adopted as the observed axial ratio ($q_{\rm b} = 1 -
\epsilon_{\rm b} = 0.78$) and position angle ($\rm PA_{\rm b} =
76\degr$) of the bulge, whereas the observed axial ratio ($q_{\rm d} =
1 - \epsilon_{\rm d} = 0.30$) and
position angle ($\rm PA_{\rm d} = 76\degr$) of the disk were
determined averaging the outer isophotes ($r>40\arcsec$). The disk
inclination ($i=77\degr$) was calculated by assuming an intrinsic
axial ratio of $q_0 = 0.21$ \citep{2008MNRAS.388.1321P}.

%

\subsection{Stellar and ionized-gas kinematics}
\label{sec:kinematics}

The stellar kinematics is measured out to more than
$50\arcsec$ on both sides along the galaxy major axis
(Fig.~\ref{fig:major}). The velocity curve of the stars is
characterized by a steep gradient in the bulge-dominated region ($|r|
< 5\arcsec$) rising to a maximum observed rotation of about 220 \kms\ at a
distance of about $15\arcsec$. At larger radii the velocity is almost
constant, or even declining in the two farthest radii measured on the
approaching SW side. The stellar velocity dispersion peaks to a
maximum of about 300 \kms\ in the center. Further out it decreases
reaching a constant value of 80 \kms\ for radii larger than $15\arcsec$.
The major-axis velocity curve of the ionized gas has a more uncertain
behaviour. It is asymmetric and extends only out to $20\arcsec$ from
the center. The gaseous component counter-rotates with respect to the
stars in the bulge-dominated region. But, the observed rotation
velocity is small ($\la 50$ \kms) and it declines to zero at larger
radii. The gas velocity dispersion is roughly constant to about 170
\kms\ for $|r|<4\arcsec$ decreasing to about 60 \kms\ outwards.

The stellar kinematics extends to about $20\arcsec$ on each side of
the nucleus along the galaxy minor axis (Fig.~\ref{fig:minor}). No
rotation is measured for the stars and their velocity dispersion falls
from a central value of 300 \kms\ to about 100 \kms\ at the last
observed radius.
The kinematics of the ionized gas was measured out to larger radii and
it is characterized by a sharp rotation. The gas velocity rises to
about 180 \kms\ in the inner $3\arcsec$. Outwards it remains constant
out to about $40\arcsec$ on the SE side, whereas it rises and declines
on the NW side. The gas velocity dispersion displays a central peak of
about 180 \kms\ and it falls to an almost constant value of 60
\kms\ for $|r| > 10\arcsec$.

The analysis of the interplay between the kinematics of the
ionized-gas and stars along both the major and minor axis of IC~5181
shows that its gaseous and stellar components are kinematically
decoupled to each other and their projected rotation axes are almost
perpendicular.

%

\subsection{Stellar populations}
\label{sec:populations}

The line-strength indices Fe5335, Fe5270, \Mgb , and \Mg2 measured
along the major axis peak in the center and decrease to an almost
constant value for $|r| > r_{\rm bd}$. Conversely, the radial profile
of \Hb\ shows a central minimum and increases outwards
(Fig.~\ref{fig:major}). This is likely to be due to differences of the
stellar populations of bulge and disk which mostly contribute to the
galaxy surface brightness within and outside $r_{\rm bd}$,
respectively (Fig.~\ref{fig:ellipse}). Indeed, the radial profiles of
the line-strength indices do not show any flat portion along the minor
axis (Fig.~\ref{fig:minor}), where the galaxy light is less
significantly contaminated by the disk contribution due to its high
inclination.

\begin{figure*}
  \centering
  \includegraphics[angle=90,width=0.48\textwidth]{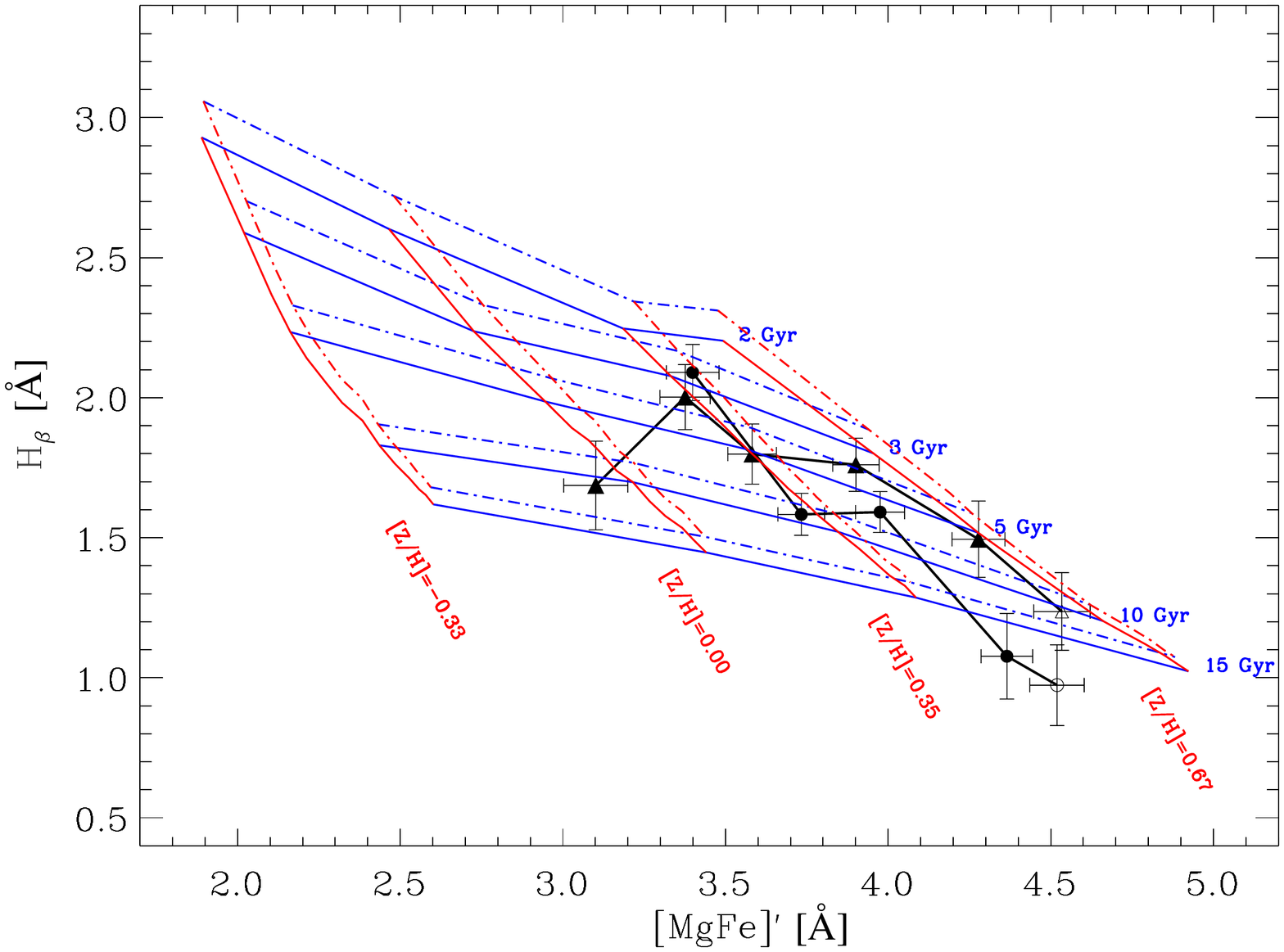}
  \includegraphics[angle=90,width=0.48\textwidth]{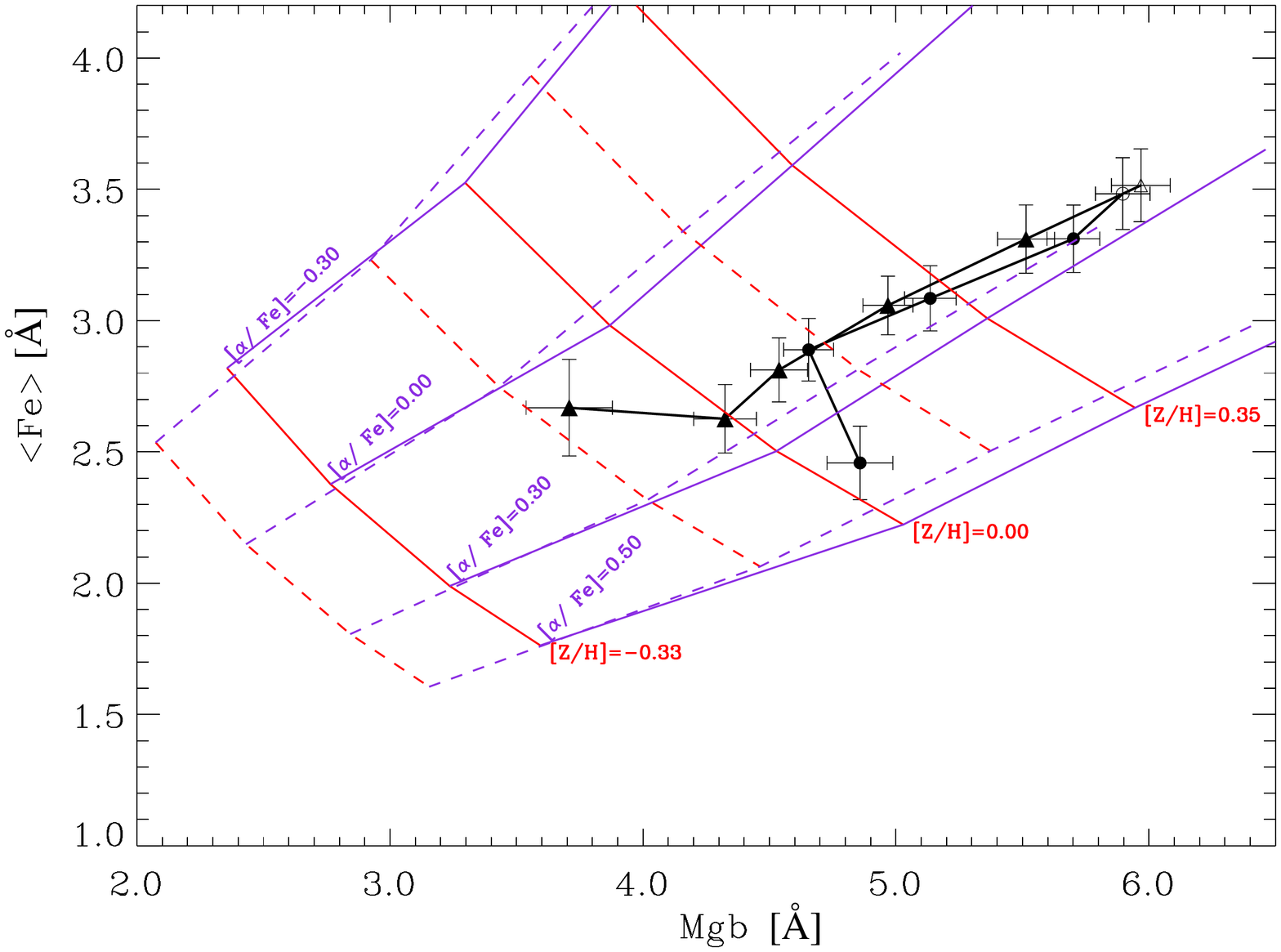}
  \caption[Distribution \Hb-\MgFe\/ and \Mgb-\Fe]{\Hb\/ and \MgFe\/
    indices (left panel) and \Fe\/ and \Mgb\/ indices (right panel)
    measured along the major (triangles) and minor axis (circles). The
    open symbols correspond to the values measured in the center of
    the galaxy and the black lines connect values measured in adjacent
    radial bins. The radial bins are centered at
    $r\,=\,0\arcsec,\,2\farcs5,\,5\farcs5,\,11\farcs3,\,23\farcs0$,
    and $41\farcs0$ along the major axis and at
    $r\,=\,0\arcsec,\,1\farcs9,\,3\farcs9,\,6\farcs5$ and $12\farcs2$ along the
    minor axis.  The grids indicate the models by
    \cite{2003MNRAS.339..897T}. In the left panel the age-metallicity
    grids are plotted with two different \aFe\ enhancements:
    \aFe$\,=\,0.0$ dex (solid lines) and \aFe$\,=\,0.5$ dex (dashed
    lines). In the right panel the \aFe\ ratio-metallicity grids are
    plotted with two different ages: 8 Gyr (solid lines) and 12 Gyr
    (dashed lines). }
   \label{fig:ssp}
\end{figure*}

Usually diagrams of different pairs of line-strength indices are
adopted for studying the properties of stellar populations. Here, we
use the line-strength indices \Hb, \Fe, and \Mgb\ measured in
Sect.~\ref{sec:data} as done in \citet{2012MNRAS.423..962M}. Balmer
lines of hydrogen are the most sensitive probes of the main sequence
turn off temperature \citep{1994ApJS...95..107W}, while the metal
lines (like \Mgb\ and \Fe) probe the temperature of the red giant
branch \citep{2000AJ....120..165T} and the abundances of these
elements. The models of \citet{2003MNRAS.339..897T} predict the values
of these line-strength indices as a function of the mean age $t$,
total metallicity \ZH , and total \aFe\ enhancement of a single stellar
population.

In Fig.~\ref{fig:ssp} we compare the prediction of the single stellar
population models and the line-strength measurements of \Hb, \Fe, and
\Mgb\ we obtained by folding their radial profiles and averaging
contiguous measurements. The radial bins were centered at
$r\,=\,0\arcsec,\,2\farcs5,\,5\farcs5,\,11\farcs3,\,23\farcs0$, and
$41\farcs0$ along the major axis and at
$r\,=\,0\arcsec,\,1\farcs9,\,3\farcs9,\,6\farcs5$ and $12\farcs2$ 
along the minor axis.

In the left panel of Fig.~\ref{fig:ssp} the values of \Hb\ and
\MgFe\ are compared with the model predictions for two stellar
populations with solar (\aFe$\,=\,0$ dex) and super-solar
\aFe\ enhancement (\aFe$\,=\,0.5$ dex), respectively. In this
parameter space, the mean age and total metallicity appear to be
almost insensitive to the variations of the \aFe\ enhancement. In the
right panel of Fig.~\ref{fig:ssp}, the values of \Mgb\ and \Fe\ are
compared with the model predictions of \citet{2003MNRAS.339..897T} for
two stellar populations with old (8 Gyr) and very old age (12 Gyr),
respectively. In this parameter space, the total metallicity and total
\aFe\ enhancement appear to be almost insensitive to the variations of
the ages.

The stellar populations in the very center of IC~5181 are very old
($t\,\ga\,12$ Gyr). The values of the age measured along both the
major and minor axis are consistent within errors.  Along the major
axis, the age decreases from 12 to 3 Gyr going from the central to the
outer ($r\,>\,r_{\rm bd}$) regions of the galaxy where the disk
contribution starts to dominate the galaxy light. This is an
indication that the stellar population of the disk is younger than
that of the bulge and it has a lower metallicity. This result is
consistent with the findings by \citet{2009MNRAS.395...28M} for a
sample of 8 spiral galaxies.  No clear differences are found for the
\aFe\ enhancement between the bulge and disk-dominated regions. As
expected for an edge-on galaxy, the disk contamination is
significantly smaller along the minor axis on account of the
inclination of the galaxy. Therefore, the corresponding age gradient
is much shallower (from 10 to 7 Gyr).
The metallicity in the bulge-dominated region has a super-solar value
in the center (\ZH$\,=\,0.58$ dex) as it results from the values
measured along both major and minor axis. It is characterized by a
negligible gradient in the inner $6\arcsec$ along the major axis
($\Delta$\ZH$\,=\,-0.06$ dex).  In the disk-dominated region the
metallicity decreases to reach a solar value at about $40\arcsec$.
The \aFe\ enhancement displays a constant super-solar value
(\aFe$\,=\,0.25$ dex) along both the major and minor axis.

The absence of radial gradients of both metallicity and
\aFe\ enhancement in the inner regions of IC~5181 ($r\,<\,r_{\rm bd}$)
suggests a bulge formation through merging events
\citep{1999ApJ...513..108B} with a very short timescale ($\Delta
t\,=\,0.5$ Gyr) for the last burst of star formation
\citep{2005ApJ...621..673T}, probably coeval with the epoch of the
merging.

%

\subsection{Ionized-gas metallicity}
\label{sec:metallicity}

To better characterize the nature of the ionized gas
we investigate its metal content and compare it to the
stellar component metallicity.

\begin{figure}
  \centering
  \includegraphics[angle=0,width=0.48\textwidth]{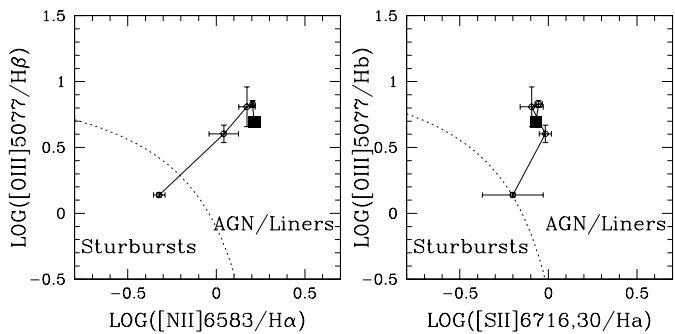}
  \caption{ The BPT diagnostic diagrams from the emission line ratios measured along 
    the minor axis of IC~5181. The radial bins are the same as in Fig.\ref{fig:ssp}.
    The larger squares mark the values measured in the central radial bin.
    The solid line connects the measurements from the central to the outermost radial bin.
    The dotted lines separate the regions occupied by starbursts and AGN/Liners and are taken from
    \citet{2001ApJ...556..121K}. }
  \label{fig:bpt}
\end{figure}

We first have to disentangle between the different sources of
ionization. To this aim we used the BPT diagnostic diagrams
\citep{1981PASP...93....5B} adopting the \oiiig/\Hb, \niig/\Ha, and
\sii/\Ha\ ratios. The emission line fluxes and their ratios were measured along the minor axis on the
same spatial bins adopted for the measurements of the line-strength indices.
Data were folded around the center and we averaged the line ratios measured at similar radii
on the two sides of the galaxy. The errors were computed as the RMS of the average. 
The results are shown in Fig.\ref{fig:bpt}.

The value of \niig/\Ha = 1.6 
found in the central regions is consistent with \citet{1986AJ.....91.1062P} and
suggests a Seyfert-type ionization mechanism.
However, the last measured points on both sides ($|r|=12$\arcsec) fall
on the limit of the starburst region of the BPT diagram implying that
in the outer part of the gaseous disk the ionization due to star formation is dominant.
Therefore, only for this region, we derived the gas-phase metallicity $12 +
\log{\rm (O/H)}$ from the empirical relation with the emission-lines
ratio parameter $R_3$ \citep{2006ApJ...652..257L} 
where the intensity
of the nebular lines \Hb\/ and \oiiipg\ was replaced with their
equivalent widths following the prescriptions by \citet{2003ApJ...599.1031K}. 
This method has the advantage of being insensitive to reddening.
Thus, the equivalent width of the \Hb\/ and \oiiipg\ emission lines
was measured in the two radial bins at r=$\pm12$\arcsec on the minor-axis spectrum.  
For each emission
line, a central bandpass covering the feature of interest and two
adjacent bandpasses, at the red and blue side, tracing the local
continuum were defined following \citet{1993PhDT.......172G} as done
in \citet{2012A&A...544A..74M}. The continuum level underlying the
emission line was estimated by interpolating a straight line in the
continuum bandpasses of the spectra. The errors associated with the
measured equivalent widths were derived from photon statistics and CCD
read-out noise, and calibrated by means of Monte Carlo simulations.

We obtained a value for the gas-phase metallicity of $12 + \log{\rm
  (O/H)}=8.6\pm0.1$ corresponding to a $\langle [Z/{\rm H}] \rangle_{\rm
  gas}\,=\,-0.08\pm0.07$ dex which was obtained as done by
\citet{2012A&A...539A.136S} and \citet{2013A&A...549A...3C}. The
latter allows a straightforward comparison with the metallicity of the
stellar populations in the disk region. The metal content of the
ionized gas is lower with respect to the stars ($\langle [Z/{\rm H}]
\rangle\,=\,0.3$ dex).  We remark that we apply the above analysis to
the measurement done at $|r|= 12$\arcsec\ were the emission lines are
mainly due to star formation, but contamination by AGN/Liners is
present. Anyway, we interpret the difference between the gas phase and
stellar component as an additional suggestion of the external origin of 
the ionized gas, which
is not related to the stars and which could have been captured by
IC~5181 during an acquisition event from the surrounding environment
or neighbouring galaxy.

\section{Discussion and conclusions}
\label{sec:conclusions}

The morphology of IC~5181 shows no remarkable peculiarity. The galaxy
disk has a high inclination with respect to the line of sight
($i=77\degr$) and it has a regular stellar body without any evidence
of warp or dust lanes.
A quantitative analysis of the photometric and spectroscopic
properties showed that IC~5181 can be considered a typical S0 galaxy
with a classical bulge according to the prescription by
\citet{2004ARA&A..42..603K}. The colour index $g-r=0.74$ and total
magnitude $M_r=-20.76$ are within the ranges measured for S0 galaxies
by \citet[][magnitudes are taken from \citet{2011ApJS..197...21H}
  and transformed to the SDSS photometric system following
  \citet{2005AJ....130..873J}]{2005AJ....129...61B}.  The bulge has a
central velocity dispersion $\sigma_0\,=\,285\,\pm\,5$ \kms\ (derived
as the mean of the values measured along the major and minor axis), an
absolute magnitude $M_{I,{\rm b}}\,=\,-20.82$, and an effective radius
$R_{\rm e} = 0.4$ kpc. It follows the Faber-Jackson correlation and
lies on the fundamental plane as the most of the bulges of S0 and SB0
galaxies \citep{2005A&A...434..109A}. In addition, it rotates as fast
as the bulges of early-type disk galaxies and it is consistent with
being an isotropic rotator (adopting $V_{\rm max}/\sigma_0 = 0.36$ and
$\epsilon_{\rm b} = 0.22$; see \citealt{2004MNRAS.354..753M} and
\citealt{2005A&A...434..109A} for a comparison). The age (10 Gyr),
metallicity (\ZH$\,=\,0.58$ dex), and overabundance (\aFe$\,=\,0.25$
dex) of the stellar populations in the bulge-dominated region match
the typical values of passively evolving S0 bulges, as derived by
\citet{2003MNRAS.339..897T}. Finally, IC~5181 fits within the scatter
the \Mgb-$\sigma$ relation for early-type galaxies by \citet[][adopting 
\Mg2$=0.346\pm0.003$]{1999MNRAS.308..833J}.

The presence of a velocity gradient along the galaxy minor axis is the
kinematic signature that ionized gas is not moving onto circular
orbits in a disk which is coplanar to that of the stars. This is
confirmed by the narrow-band imaging showing the distribution of the
gas in the innermost regions of IC~5181.  Moreover, the short
dynamical time (few Myr) we derived for the galaxy regions where we
observe the velocity gradient and the regular velocity field of the
ionized gas implies that it is already settled in an equilibrium
configuration \citep[e.g.,][]{1985ApJ...292L..51B}.  Since the
intrinsic shape of bulges is triaxial \citep{1991ApJ...374L..13B, 
  2008A&A...478..353M, 2010A&A...521A..71M} and two equilibrium planes
are allowed for the gaseous component
\citep[e.g.,][]{1989ApJ...343..617D}, we can explain the observed
kinematics of IC~5181 as due to ionized gas settled on the equilibrium
plane perpendicular to the long axis of a triaxial bulge. The ionized
gas is in orthogonal rotation with respect to the galaxy disk and we
measure zero velocity (or at least a shallower velocity gradient)
along the disk major axis and a sharp velocity gradient along the disk
minor axis.

To constrain the orientation of the triaxial bulge we derived the
orientation of the gaseous disk with respect to the line of sight. To
this aim we compared the circular velocity $v_{\rm circ}$ obtained
from the major-axis stellar kinematics with the rotation velocity of
the ionized gas $v_{\rm gas}$ measured along the minor axis. The
comparison is restricted where the pressure support of the gas is
small and therefore the gas velocity is supposed to trace the circular
velocity \citep{1995ApJ...448L..13B, 2001MNRAS.323..188P,
  2010ApJ...721..547D}.  All the observed velocities were subtracted
of systemic velocity and folded around the galaxy center. For radii
larger than $15\arcsec$ we obtained a nearly flat circular velocity
curve with $v_{\rm circ} = 305\pm10$ \kms\ by applying the asymmetric
drift correction to the major-axis stellar kinematics as done by
\citet{2003MNRAS.338..465A} after dealing with the effects of
integration along the line of sight following
\citet{1999AJ....117.2666N}. The ionized-gas rotation curve was
derived from the data measured along the minor axis at radii larger
than $8\arcsec$ where $\sigma_{\rm gas}<60$ \kms. It is flat and shows
a constant value $v_{\rm gas} = 201\pm6$ \kms. We deprojected $v_{\rm
  gas}$ into $v_{\rm circ}$ as done by \citet{2012AJ....144...78W},
and assuming that the gaseous component is moving onto circular orbits
in an infinitesimally thin disk perpendicular to the galaxy disk. For
the gaseous disk we derived the inclination ($i_{\rm gas} =
41\degr\pm3\degr$) and position angle of its angular momentum ($\rm
PA_{\rm gas} = 62\degr\pm2\degr$). The small difference ($\rm \Delta
PA=14\degr$) between the position angles of the disk major axis and
bulge long axis (which coincides with $\rm PA_{\rm gas}$) prevented us
to successfully model the surface brightness distribution of the
triaxial bulge and properly derive its intrinsic shape and
contribution to the total light of the galaxy \citep[e.g., as done
  by][]{2010A&A...521A..71M, 2012MNRAS.423L..79C}.

As an alternative, the gas is moving on anomalous orbits in a
triaxial bulge (or bar) that is tumbling about its short axis
\citep[see][for details]{1982MNRAS.201..303V, 1993A&A...268...65F}.
This is the case of the lenticular galaxy NGC~128, which hosts a
counter-rotating disk of ionized gas \citep{1997A&A...318L..39E,
  2012MNRAS.422.1083C}. Like for IC~5181, the orientation of the
ionized gas is not aligned with the stellar disk. This is due to the
presence of a tumbling triaxial bar associated with the peanut
morphology. In IC~5181 we do not observe gas in retrograde motion
relative to stars at large radii from the galaxy center, where the
anomalous gas orbits are expected to be highly inclined with respect
to the figure rotation axis since the triaxial bulge (or bar) is seen
close to end-on.  In this scenario the gas is settled into a stable
configuration forming a strongly warped disk, whose innermost portion
corresponds to an inner polar disk \citep{2003A&A...408..873C,
  2004AJ....127.2641S}.  For IC~5181 there is no evidence of stars
associated to the orthogonally-rotating gas neither from the observed
kinematics nor from the analysis of the stellar populations.

Disentangling between the two alternative geometric configurations
with either the innermost or all the ionized-gas component in
orthogonal rotation with respect to the stellar disk requires the
analysis of the full velocity field of the galaxy measured with
integral-field spectroscopy \citep{2007A&A...465..777C}. Nevertheless,
the kinematical decoupling between the gaseous and stellar components
of IC~5181 suggests the occurrence of an accretion event or merging
\citep{1999IAUS..186..149B}.  Therefore, it is straightforward to
explain the existence of the orthogonally rotating gas in IC~5181 as
the end result of the acquisition of external gas by the pre-existing
galaxy.  The nearby environment of IC~5181 shows no strong evidence
for such an event. IC~5181 is not interacting with its closest
companion, NGC~7232A, and it is one of the few galaxies of the
NGC~7213 group to be undetected in \hi . However, all these arguments
do not exclude an external origin of the ionized-gas component of
IC~5181. \citet{2001A&A...380..471B} proved that the environment of
galaxies that experienced past gas accretion do not appear
statistically different from those of normal galaxies. In addition, 
the polar orientation of the ionized gas and its possible low metal content 
fits well with the scenario proposed for the formation of polar ring
galaxies, where the gaseous ring is formed by the accretion of
material from a cosmic filament \citep{2006ApJ...636L..25M,
  2008ApJ...689..678B}. 

\begin{acknowledgements}
We thank Michela Mapelli and Marilena Spavone for useful discussions. 
This work was supported by Padua University through grants
60A02-1283/10, 60A02-5052/11, and 60A02-4807/12. MC and LM acknowledge
financial support from Padua University grant CPDR115539/11 and
CPS0204, respectively.
\end{acknowledgements}

\bibliographystyle{aa} 

\begin{thebibliography}{87}
\expandafter\ifx\csname natexlab\endcsname\relax\def\natexlab#1{#1}\fi

\bibitem[{{Aguerri} {et~al.}(2003){Aguerri}, {Debattista}, \&
  {Corsini}}]{2003MNRAS.338..465A}
{Aguerri}, J.~A.~L., {Debattista}, V.~P., \& {Corsini}, E.~M. 2003, \mnras,
  338, 465

\bibitem[{{Aguerri} {et~al.}(2005){Aguerri}, {Elias-Rosa}, {Corsini}, \&
  {Mu{\~n}oz-Tu{\~n}{\'o}n}}]{2005A&A...434..109A}
{Aguerri}, J.~A.~L., {Elias-Rosa}, N., {Corsini}, E.~M., \&
  {Mu{\~n}oz-Tu{\~n}{\'o}n}, C. 2005, \aap, 434, 109

\bibitem[{{Athanassoula}(2003)}]{2003MNRAS.341.1179A}
{Athanassoula}, E. 2003, \mnras, 341, 1179

\bibitem[{{Athanassoula}(2005)}]{2005MNRAS.358.1477A}
{Athanassoula}, E. 2005, \mnras, 358, 1477

\bibitem[Baldwin et al.(1981)]{1981PASP...93....5B} Baldwin, J.~A., 
Phillips, M.~M., \& Terlevich, R.\ 1981, \pasp, 93, 5 


\bibitem[{{Bekki} \& {Shioya}(1999)}]{1999ApJ...513..108B}
{Bekki}, K. \& {Shioya}, Y. 1999, \apj, 513, 108

\bibitem[{{Bernardi} {et~al.}(2005){Bernardi}, {Sheth}, {Nichol}, {Schneider},
  \& {Brinkmann}}]{2005AJ....129...61B}
{Bernardi}, M., {Sheth}, R.~K., {Nichol}, R.~C., {Schneider}, D.~P., \&
  {Brinkmann}, J. 2005, \aj, 129, 61

\bibitem[{{Bertola} {et~al.}(1992){Bertola}, {Buson}, \&
  {Zeilinger}}]{1992ApJ...401L..79B}
{Bertola}, F., {Buson}, L.~M., \& {Zeilinger}, W.~W. 1992, \apjl, 401, L79

\bibitem[{{Bertola} {et~al.}(1996){Bertola}, {Cinzano}, {Corsini}, {Pizzella},
  {Persic}, \& {Salucci}}]{1996ApJ...458L..67B}
{Bertola}, F., {Cinzano}, P., {Corsini}, E.~M., {et~al.} 1996, \apjl, 458, L67

\bibitem[{{Bertola} {et~al.}(1995){Bertola}, {Cinzano}, {Corsini}, {Rix}, \&
  {Zeilinger}}]{1995ApJ...448L..13B}
{Bertola}, F., {Cinzano}, P., {Corsini}, E.~M., {Rix}, H.-W., \& {Zeilinger},
  W.~W. 1995, \apjl, 448, L13

\bibitem[{{Bertola} \& {Corsini}(1999)}]{1999IAUS..186..149B}
{Bertola}, F. \& {Corsini}, E.~M. 1999, in IAU Symposium, Vol. 186, Galaxy
  Interactions at Low and High Redshift, ed. J.~E. {Barnes} \& D.~B. {Sanders},
 (Kluwer, Dordrecht), 149

\bibitem[{{Bertola} {et~al.}(1999){Bertola}, {Corsini}, {Vega Beltr{\'a}n},
  {Pizzella}, {Sarzi}, {Cappellari}, \& {Funes}}]{1999ApJ...519L.127B}
{Bertola}, F., {Corsini}, E.~M., {Vega Beltr{\'a}n}, J.~C., {et~al.} 1999,
  \apjl, 519, L127

\bibitem[{{Bertola} {et~al.}(1985){Bertola}, {Galletta}, \&
  {Zeilinger}}]{1985ApJ...292L..51B}
{Bertola}, F., {Galletta}, G., \& {Zeilinger}, W.~W. 1985, \apjl, 292, L51

\bibitem[{{Bertola} {et~al.}(1991){Bertola}, {Vietri}, \&
  {Zeilinger}}]{1991ApJ...374L..13B}
{Bertola}, F., {Vietri}, M., \& {Zeilinger}, W.~W. 1991, \apjl, 374, L13

\bibitem[{{Bettoni} {et~al.}(2001){Bettoni}, {Falomo}, {Fasano}, {Govoni},
  {Salvo}, \& {Scarpa}}]{2001A&A...380..471B}
{Bettoni}, D., {Falomo}, R., {Fasano}, G., {et~al.} 2001, \aap, 380, 471

\bibitem[{{Brook} {et~al.}(2008){Brook}, {Governato}, {Quinn}, {Wadsley},
  {Brooks}, {Willman}, {Stilp}, \& {Jonsson}}]{2008ApJ...689..678B}
{Brook}, C.~B., {Governato}, F., {Quinn}, T., {et~al.} 2008, \apj, 689, 678

\bibitem[{{Bureau} {et~al.}(2006){Bureau}, {Aronica}, {Athanassoula},
  {Dettmar}, {Bosma}, \& {Freeman}}]{2006MNRAS.370..753B}
{Bureau}, M., {Aronica}, G., {Athanassoula}, E., {et~al.} 2006, \mnras, 370,
  753

\bibitem[{{Burki} {et~al.}(1995){Burki}, {Rufener}, {Burnet}, {Richard},
  {Blecha}, \& {Bratschi}}]{1995Msngr..80...34B}
{Burki}, G., {Rufener}, F., {Burnet}, M., {et~al.} 1995, The Messenger, 80, 34

\bibitem[{{Chung} {et~al.}(2012){Chung}, {Bureau}, {van Gorkom}, \&
  {Koribalski}}]{2012MNRAS.422.1083C}
{Chung}, A., {Bureau}, M., {van Gorkom}, J.~H., \& {Koribalski}, B. 2012,
  \mnras, 422, 1083

\bibitem[{{Ciri} {et~al.}(1995){Ciri}, {Bettoni}, \&
  {Galletta}}]{1995Natur.375..661C}
{Ciri}, R., {Bettoni}, D., \& {Galletta}, G. 1995, \nat, 375, 661

\bibitem[{{Coccato} {et~al.}(2007){Coccato}, {Corsini}, {Pizzella}, \&
  {Bertola}}]{2007A&A...465..777C}
{Coccato}, L., {Corsini}, E.~M., {Pizzella}, A., \& {Bertola}, F. 2007, \aap,
  465, 777

\bibitem[{{Coccato} {et~al.}(2013){Coccato}, {Morelli}, {Pizzella}, {Corsini},
  {Buson}, \& {Dalla Bont{\`a}}}]{2013A&A...549A...3C}
{Coccato}, L., {Morelli}, L., {Pizzella}, A., {et~al.} 2013, \aap, 549, A3

\bibitem[{{Combes} {et~al.}(2013){Combes}, {Moiseev}, \&
  {Reshetnikov}}]{2013arXiv1302.7273C}
{Combes}, F., {Moiseev}, A., \& {Reshetnikov}, V. 2013, \aap, 554, A11

\bibitem[{{Corsini} {et~al.}(2012){Corsini}, {M{\'e}ndez-Abreu}, {Pastorello},
  {Dalla Bont{\`a}}, {Morelli}, {Beifiori}, {Pizzella}, \&
  {Bertola}}]{2012MNRAS.423L..79C}
{Corsini}, E.~M., {M{\'e}ndez-Abreu}, J., {Pastorello}, N., {et~al.} 2012,
  \mnras, 423, L79

\bibitem[{{Corsini} {et~al.}(2003){Corsini}, {Pizzella}, {Coccato}, \&
  {Bertola}}]{2003A&A...408..873C}
{Corsini}, E.~M., {Pizzella}, A., {Coccato}, L., \& {Bertola}, F. 2003, \aap,
  408, 873

\bibitem[{{Dalcanton} \& {Stilp}(2010)}]{2010ApJ...721..547D}
{Dalcanton}, J.~J. \& {Stilp}, A.~M. 2010, \apj, 721, 547

\bibitem[{{Davis} {et~al.}(2011){Davis}, {Alatalo}, {Sarzi}, {Bureau}, {Young},
  {Blitz}, {Serra}, {Crocker}, {Krajnovi{\'c}}, {McDermid}, {Bois}, {Bournaud},
  {Cappellari}, {Davies}, {Duc}, {de Zeeuw}, {Emsellem}, {Khochfar},
  {Kuntschner}, {Lablanche}, {Morganti}, {Naab}, {Oosterloo}, {Scott}, \&
  {Weijmans}}]{2011MNRAS.417..882D}
{Davis}, T.~A., {Alatalo}, K., {Sarzi}, M., {et~al.} 2011, \mnras, 417, 882

\bibitem[{{de Vaucouleurs} {et~al.}(1991){de Vaucouleurs}, {de Vaucouleurs},
  {Corwin}, {Buta}, {Paturel}, \& {Fouqu{\'e}}}]{1991rc3..book.....D}
{de Vaucouleurs}, G., {de Vaucouleurs}, A., {Corwin}, Jr., H.~G., {et~al.}
  1991, {Third Reference Catalogue of Bright Galaxies Vol. 1-3} (New York: Springer)

\bibitem[{{de Zeeuw} \& {Franx}(1989)}]{1989ApJ...343..617D}
{de Zeeuw}, T. \& {Franx}, M. 1989, \apj, 343, 617

\bibitem[{{Emsellem} \& {Arsenault}(1997)}]{1997A&A...318L..39E}
{Emsellem}, E. \& {Arsenault}, R. 1997, \aap, 318, L39

\bibitem[{{Emsellem} {et~al.}(2006){Emsellem}, {Fathi}, {Wozniak}, {Ferruit},
  {Mundell}, \& {Schinnerer}}]{2006MNRAS.365..367E}
{Emsellem}, E., {Fathi}, K., {Wozniak}, H., {et~al.} 2006, \mnras, 365, 367

\bibitem[{{Friedli} \& {Benz}(1993)}]{1993A&A...268...65F}
{Friedli}, D. \& {Benz}, W. 1993, \aap, 268, 65

\bibitem[{{Garcia}(1993)}]{1993A&AS..100...47G}
{Garcia}, A.~M. 1993, \aaps, 100, 47

\bibitem[{{Gerhard}(1993)}]{1993MNRAS.265..213G}
{Gerhard}, O.~E. 1993, \mnras, 265, 213

\bibitem[{{Gonz{\'a}lez}(1993)}]{1993PhDT.......172G}
{Gonz{\'a}lez}, J.~J. 1993, PhD thesis, University of California

\bibitem[{{Gorgas} {et~al.}(1990){Gorgas}, {Efstathiou}, \& {Aragon
  Salamanca}}]{1990MNRAS.245..217G}
{Gorgas}, J., {Efstathiou}, G., \& {Aragon Salamanca}, A. 1990, \mnras, 245,
  217

\bibitem[{{Haynes} {et~al.}(1984){Haynes}, {Giovanelli}, \&
  {Chincarini}}]{1984ARA&A..22..445H}
{Haynes}, M.~P., {Giovanelli}, R., \& {Chincarini}, G.~L. 1984, \araa, 22, 445

\bibitem[{{Ho} {et~al.}(2011){Ho}, {Li}, {Barth}, {Seigar}, \&
  {Peng}}]{2011ApJS..197...21H}
{Ho}, L.~C., {Li}, Z.-Y., {Barth}, A.~J., {Seigar}, M.~S., \& {Peng}, C.~Y.
  2011, \apjs, 197, 21

\bibitem[{{Iodice} {et~al.}(2002){Iodice}, {Arnaboldi}, {Sparke}, {Gallagher},
  \& {Freeman}}]{2002A&A...391..103I}
{Iodice}, E., {Arnaboldi}, M., {Sparke}, L.~S., {Gallagher}, J.~S., \&
  {Freeman}, K.~C. 2002, \aap, 391, 103

\bibitem[{{Jester} {et~al.}(2005){Jester}, {Schneider}, {Richards}, {Green},
  {Schmidt}, {Hall}, {Strauss}, {Vanden Berk}, {Stoughton}, {Gunn},
  {Brinkmann}, {Kent}, {Smith}, {Tucker}, \& {Yanny}}]{2005AJ....130..873J}
{Jester}, S., {Schneider}, D.~P., {Richards}, G.~T., {et~al.} 2005, \aj, 130,
  873

\bibitem[{{J{\o}rgensen} {et~al.}(1999){J{\o}rgensen}, {Franx}, {Hjorth}, \&
  {van Dokkum}}]{1999MNRAS.308..833J}
{J{\o}rgensen}, I., {Franx}, M., {Hjorth}, J., \& {van Dokkum}, P.~G. 1999,
  \mnras, 308, 833

\bibitem[{{Kannappan} \& {Fabricant}(2001)}]{2001AJ....121..140K}
{Kannappan}, S.~J. \& {Fabricant}, D.~G. 2001, \aj, 121, 140

\bibitem[{{Katkov} {et~al.}(2013){Katkov}, {Sil'chenko}, \&
  {Afanasiev}}]{2013ApJ...769..105K}
{Katkov}, I.~Y., {Sil'chenko}, O.~K., \& {Afanasiev}, V.~L. 2013, \apj, 769,
  105

\bibitem[{{Kewley} {et~al.}(2001){Dopita}, {Sutherland}, {Heisler}, \&
    {Trevena}}]{2001ApJ...556..121K} {Kewley}, L.~J., {Dopita}, M.~A.,
  {Sutherland}, R.~S., {Heisler}, C.~A., \& {Trevena}, J.\ 2001, \apj,
  556, 121

\bibitem[{{Kobulnicky} \& {Phillips}(2003)}]{2003ApJ...599.1031K}
{Kobulnicky}, H.~A. \& {Phillips}, A.~C. 2003, \apj, 599, 1031

\bibitem[{{Kormendy}(1977)}]{1977ApJ...217..406K}
{Kormendy}, J. 1977, \apj, 217, 406

\bibitem[{{Kormendy} \& {Kennicutt}(2004)}]{2004ARA&A..42..603K}
{Kormendy}, J. \& {Kennicutt}, Jr., R.~C. 2004, \araa, 42, 603

\bibitem[{{Krajnovi{\'c}} \& {Jaffe}(2004)}]{2004A&A...428..877K}
{Krajnovi{\'c}}, D. \& {Jaffe}, W. 2004, \aap, 428, 877

\bibitem[{{Kuijken} {et~al.}(1996){Kuijken}, {Fisher}, \&
  {Merrifield}}]{1996MNRAS.283..543K}
{Kuijken}, K., {Fisher}, D., \& {Merrifield}, M.~R. 1996, \mnras, 283, 543

\bibitem[{{Liang} {et~al.}(2006){Liang}, {Yin}, {Hammer}, {Deng}, {Flores}, \&
  {Zhang}}]{2006ApJ...652..257L}
{Liang}, Y.~C., {Yin}, S.~Y., {Hammer}, F., {et~al.} 2006, \apj, 652, 257

\bibitem[{{MacArthur} {et~al.}(2009){MacArthur}, {Gonz{\'a}lez}, \&
  {Courteau}}]{2009MNRAS.395...28M}
{MacArthur}, L.~A., {Gonz{\'a}lez}, J.~J., \& {Courteau}, S. 2009, \mnras, 395,
  28

\bibitem[{{Macci{\`o}} {et~al.}(2006){Macci{\`o}}, {Moore}, \&
  {Stadel}}]{2006ApJ...636L..25M}
{Macci{\`o}}, A.~V., {Moore}, B., \& {Stadel}, J. 2006, \apjl, 636, L25

\bibitem[{{Maia} {et~al.}(1989){Maia}, {da Costa}, \&
  {Latham}}]{1989ApJS...69..809M}
{Maia}, M.~A.~G., {da Costa}, L.~N., \& {Latham}, D.~W. 1989, \apjs, 69, 809

\bibitem[{{Markwardt}(2009)}]{2009ASPC..411..251M}
{Markwardt}, C.~B. 2009, in Astronomical Society of the Pacific Conference
  Series, Vol. 411, Astronomical Data Analysis Software and Systems XVIII, ed.
  D.~A. {Bohlender}, D.~{Durand}, \& P.~{Dowler},  (San Francisco, CA: ASP), 251 

\bibitem[{{Matthews} \& {de Grijs}(2004)}]{2004AJ....128..137M}
{Matthews}, L.~D. \& {de Grijs}, R. 2004, \aj, 128, 137

\bibitem[{{M{\'e}ndez-Abreu} {et~al.}(2008){M{\'e}ndez-Abreu}, {Aguerri},
  {Corsini}, \& {Simonneau}}]{2008A&A...478..353M}
{M{\'e}ndez-Abreu}, J., {Aguerri}, J.~A.~L., {Corsini}, E.~M., \& {Simonneau},
  E. 2008, \aap, 478, 353

\bibitem[{{M{\'e}ndez-Abreu} {et~al.}(2010){M{\'e}ndez-Abreu}, {Simonneau},
  {Aguerri}, \& {Corsini}}]{2010A&A...521A..71M}
{M{\'e}ndez-Abreu}, J., {Simonneau}, E., {Aguerri}, J.~A.~L., \& {Corsini},
  E.~M. 2010, \aap, 521, A71

\bibitem[{{Moiseev} {et~al.}(2011){Moiseev}, {Smirnova}, {Smirnova}, \&
  {Reshetnikov}}]{2011MNRAS.418..244M}
{Moiseev}, A.~V., {Smirnova}, K.~I., {Smirnova}, A.~A., \& {Reshetnikov}, V.~P.
  2011, \mnras, 418, 244

\bibitem[{{Morelli} {et~al.}(2012{\natexlab{a}}){Morelli}, {Calvi}, {Cardullo},
  {Pizzella}, {Corsini}, \& {Dalla Bont{\`a}}}]{2012A&A...544A..74M}
{Morelli}, L., {Calvi}, V., {Cardullo}, A., {et~al.} 2012{\natexlab{a}}, \aap,
  544, A74

\bibitem[{{Morelli} {et~al.}(2012{\natexlab{b}}){Morelli}, {Corsini},
  {Pizzella}, {Dalla Bont{\`a}}, {Coccato}, {M{\'e}ndez-Abreu}, \&
  {Cesetti}}]{2012MNRAS.423..962M}
{Morelli}, L., {Corsini}, E.~M., {Pizzella}, A., {et~al.} 2012{\natexlab{b}},
  \mnras, 423, 962

\bibitem[{{Morelli} {et~al.}(2004){Morelli}, {Halliday}, {Corsini}, {Pizzella},
  {Thomas}, {Saglia}, {Davies}, {Bender}, {Birkinshaw}, \&
  {Bertola}}]{2004MNRAS.354..753M}
{Morelli}, L., {Halliday}, C., {Corsini}, E.~M., {et~al.} 2004, \mnras, 354,
  753

\bibitem[{{Neistein} {et~al.}(1999){Neistein}, {Maoz}, {Rix}, \&
  {Tonry}}]{1999AJ....117.2666N}
{Neistein}, E., {Maoz}, D., {Rix}, H.-W., \& {Tonry}, J.~L. 1999, \aj, 117,
  2666

\bibitem[{{Osterbrock}(1989)}]{1989agna.book.....O}
{Osterbrock}, D.~E. 1989, {Astrophysics of gaseous nebulae and active galactic
  nuclei} (Mill Valley, CA: University Science Books)

\bibitem[{{Padilla} \& {Strauss}(2008)}]{2008MNRAS.388.1321P}
{Padilla}, N.~D. \& {Strauss}, M.~A. 2008, \mnras, 388, 1321

\bibitem[{{Peng} {et~al.}(2010){Peng}, {Ho}, {Impey}, \&
  {Rix}}]{2010AJ....139.2097P}
{Peng}, C.~Y., {Ho}, L.~C., {Impey}, C.~D., \& {Rix}, H.-W. 2010, \aj, 139,
  2097

\bibitem[{{Phillips} {et~al.}(1986){Phillips}, {Jenkins}, 
{Dopita}, {Sadler}, \& {Binette}}]{1986AJ.....91.1062P} 
{Phillips}, M.~M., {Jenkins}, C.~R., {Dopita}, M.~A., {Sadler}, E.~M., 
\& {Binette}, L.\ 1986, \aj, 91, 1062 

\bibitem[{{Pignatelli} {et~al.}(2001){Pignatelli}, {Corsini}, {Vega
  Beltr{\'a}n}, {Scarlata}, {Pizzella}, {Funes}, {Zeilinger}, {Beckman}, \&
  {Bertola}}]{2001MNRAS.323..188P}
{Pignatelli}, E., {Corsini}, E.~M., {Vega Beltr{\'a}n}, J.~C., {et~al.} 2001,
  \mnras, 323, 188

\bibitem[{{Pizzella} {et~al.}(1997){Pizzella}, {Amico}, {Bertola}, {Buson},
  {Danziger}, {Dejonghe}, {Sadler}, {Saglia}, {de Zeeuw}, \&
  {Zeilinger}}]{1997A&A...323..349P}
{Pizzella}, A., {Amico}, P., {Bertola}, F., {et~al.} 1997, \aap, 323, 349

\bibitem[{{Pizzella} {et~al.}(2008){Pizzella}, {Corsini}, {Sarzi}, {Magorrian},
  {M{\'e}ndez-Abreu}, {Coccato}, {Morelli}, \& {Bertola}}]{2008MNRAS.387.1099P}
{Pizzella}, A., {Corsini}, E.~M., {Sarzi}, M., {et~al.} 2008, \mnras, 387, 1099

\bibitem[{{Pizzella} {et~al.}(2004){Pizzella}, {Corsini}, {Vega Beltr{\'a}n},
  \& {Bertola}}]{2004A&A...424..447P}
{Pizzella}, A., {Corsini}, E.~M., {Vega Beltr{\'a}n}, J.~C., \& {Bertola}, F.
  2004, \aap, 424, 447

\bibitem[{{Prieto} {et~al.}(2001){Prieto}, {Aguerri}, {Varela}, \&
  {Mu{\~n}oz-Tu{\~n}{\'o}n}}]{2001A&A...367..405P}
{Prieto}, M., {Aguerri}, J.~A.~L., {Varela}, A.~M., \&
  {Mu{\~n}oz-Tu{\~n}{\'o}n}, C. 2001, \aap, 367, 405

\bibitem[{{Rubin}(1994)}]{1994AJ....108..456R}
{Rubin}, V.~C. 1994, \aj, 108, 456

\bibitem[{{Rubin} {et~al.}(1992){Rubin}, {Graham}, \&
  {Kenney}}]{1992ApJ...394L...9R}
{Rubin}, V.~C., {Graham}, J.~A., \& {Kenney}, J.~D.~P. 1992, \apjl, 394, L9

\bibitem[{{S{\'a}nchez-Bl{\'a}zquez} {et~al.}(2006){S{\'a}nchez-Bl{\'a}zquez},
  {Peletier}, {Jim{\'e}nez-Vicente}, {Cardiel}, {Cenarro},
  {Falc{\'o}n-Barroso}, {Gorgas}, {Selam}, \& {Vazdekis}}]{2006MNRAS.371..703S}
{S{\'a}nchez-Bl{\'a}zquez}, P., {Peletier}, R.~F., {Jim{\'e}nez-Vicente}, J.,
  {et~al.} 2006, \mnras, 371, 703

\bibitem[{{Sandage} \& {Bedke}(1994)}]{1994cag..book.....S}
{Sandage}, A. \& {Bedke}, J. 1994, {The Carnegie Atlas of Galaxies. Volumes I,
  II.}, (Washington, DC: Carnegie Inst. of Washington) 

\bibitem[{{Sarzi} {et~al.}(2006){Sarzi}, {Falc{\'o}n-Barroso}, {Davies},
  {Bacon}, {Bureau}, {Cappellari}, {de Zeeuw}, {Emsellem}, {Fathi},
  {Krajnovi{\'c}}, {Kuntschner}, {McDermid}, \&
  {Peletier}}]{2006MNRAS.366.1151S}
{Sarzi}, M., {Falc{\'o}n-Barroso}, J., {Davies}, R.~L., {et~al.} 2006, \mnras,
  366, 1151

\bibitem[{{Schweizer}(1998)}]{1998giis.conf..105S}
{Schweizer}, F. 1998, in Saas-Fee Advanced Course 26: Galaxies: Interactions
  and Induced Star Formation, ed. R.~C. {Kennicutt}, Jr., F.~{Schweizer}, J.~E.
  {Barnes}, D.~{Friedli}, L.~{Martinet}, \& D.~{Pfenniger}, (New York: Springer), 105

\bibitem[{{Sil'chenko} \& {Afanasiev}(2004)}]{2004AJ....127.2641S}
{Sil'chenko}, O.~K. \& {Afanasiev}, V.~L. 2004, \aj, 127, 2641

\bibitem[{{Sil'chenko} {et~al.}(2009){Sil'chenko}, {Moiseev}, \&
  {Afanasiev}}]{2009ApJ...694.1550S}
{Sil'chenko}, O.~K., {Moiseev}, A.~V., \& {Afanasiev}, V.~L. 2009, \apj, 694,
  1550

\bibitem[{{Sommariva} {et~al.}(2012){Sommariva}, {Mannucci}, {Cresci},
  {Maiolino}, {Marconi}, {Nagao}, {Baroni}, \& {Grazian}}]{2012A&A...539A.136S}
{Sommariva}, V., {Mannucci}, F., {Cresci}, G., {et~al.} 2012, \aap, 539, A136

\bibitem[{{Spavone} {et~al.}(2010){Spavone}, {Iodice}, {Arnaboldi}, {Gerhard},
  {Saglia}, \& {Longo}}]{2010ApJ...714.1081S}
{Spavone}, M., {Iodice}, E., {Arnaboldi}, M., {et~al.} 2010, \apj, 714, 1081

\bibitem[{{Thomas} {et~al.}(2003){Thomas}, {Maraston}, \&
  {Bender}}]{2003MNRAS.339..897T}
{Thomas}, D., {Maraston}, C., \& {Bender}, R. 2003, \mnras, 339, 897

\bibitem[{{Thomas} {et~al.}(2005){Thomas}, {Maraston}, {Bender}, \& {Mendes de
  Oliveira}}]{2005ApJ...621..673T}
{Thomas}, D., {Maraston}, C., {Bender}, R., \& {Mendes de Oliveira}, C. 2005,
  \apj, 621, 673

\bibitem[{{Trager} {et~al.}(2000){Trager}, {Faber}, {Worthey}, \&
  {Gonz{\'a}lez}}]{2000AJ....120..165T}
{Trager}, S.~C., {Faber}, S.~M., {Worthey}, G., \& {Gonz{\'a}lez}, J.~J. 2000,
  \aj, 120, 165

\bibitem[{{Tully}(1988)}]{1988ngc..book.....T}
{Tully}, R.~B. 1988, {Nearby galaxies catalog}, (Cambridge: Cambridge Univ. Press)

\bibitem[{{van Albada} \& {Sanders}(1982)}]{1982MNRAS.201..303V}
{van Albada}, T.~S. \& {Sanders}, R.~H. 1982, \mnras, 201, 303

\bibitem[{{van der Marel} \& {Franx}(1993)}]{1993ApJ...407..525V}
{van der Marel}, R.~P. \& {Franx}, M. 1993, \apj, 407, 525

\bibitem[{{Wegner} {et~al.}(2012){Wegner}, {Corsini}, {Thomas}, {Saglia},
  {Bender}, \& {Pu}}]{2012AJ....144...78W}
{Wegner}, G.~A., {Corsini}, E.~M., {Thomas}, J., {et~al.} 2012, \aj, 144, 78

\bibitem[{{Whitmore} {et~al.}(1990){Whitmore}, {Lucas}, {McElroy},
  {Steiman-Cameron}, {Sackett}, \& {Olling}}]{1990AJ....100.1489W}
{Whitmore}, B.~C., {Lucas}, R.~A., {McElroy}, D.~B., {et~al.} 1990, \aj, 100,
  1489

\bibitem[{{Worthey}(1994)}]{1994ApJS...95..107W}
{Worthey}, G. 1994, \apjs, 95, 107

\end{thebibliography}

\newpage
\onecolumn
\begin{center}
{\huge On-line tables}
\end{center}

\begin{longtable}{rrrrrrrcrc}
\caption{Stellar kinematics along the major (PA=76\degr ) and minor 
(PA=164\degr ) axis of IC~5181.}\\
\hline
\multicolumn{1}{c}{PA}       &
\multicolumn{1}{c}{r}        & 
\multicolumn{2}{c}{V$\pm$dV} &
\multicolumn{2}{c}{$\sigma\pm d\sigma$} &
\multicolumn{2}{c}{h$_3\pm$dh$_3$} &
\multicolumn{2}{c}{h$_4\pm$dh$_4$} \\

\multicolumn{1}{c}{\degr}   &    
\multicolumn{1}{c}{\arcsec} &    
\multicolumn{2}{c}{\kms}    &
\multicolumn{2}{c}{\kms}    &
\multicolumn{2}{c}{}        &
\multicolumn{2}{c}{} \\

\hline
 76 & -50.0 &  213 &  14 & 153 &  19 &  0.103 & 0.056 & -0.022 & 0.096\\
 76 & -44.1 &  210 &  12 &  78 &  30 &  0.038 & 0.107 & -0.017 & 0.221\\
 76 & -39.6 &  210 &   8 &  79 &  21 & -0.023 & 0.066 &  0.023 & 0.162\\
 76 & -36.0 &  219 &   9 &  82 &  14 & -0.075 & 0.080 &  0.056 & 0.111\\
 76 & -33.2 &  232 &  11 &  82 &  11 & -0.002 & 0.102 &  0.041 & 0.065\\
 76 & -30.7 &  226 &  10 &  94 &  11 &  0.025 & 0.081 &  0.076 & 0.064\\
 76 & -28.7 &  233 &   7 &  78 &  17 & -0.090 & 0.058 &  0.025 & 0.135\\
 76 & -27.0 &  228 &  10 &  86 &  13 & -0.049 & 0.088 &  0.068 & 0.086\\
 76 & -25.4 &  236 &  14 &  56 &  20 & -0.012 & 0.176 &  0.025 & 0.191\\
 76 & -23.7 &  236 &   8 & 103 &  13 & -0.110 & 0.052 &  0.032 & 0.092\\
 76 & -22.1 &  204 &   8 & 110 &  10 &  0.000 & 0.056 &  0.104 & 0.056\\
 76 & -20.5 &  198 &   8 &  95 &  10 &  0.026 & 0.063 &  0.061 & 0.062\\
 76 & -18.8 &  213 &   8 &  90 &  10 & -0.023 & 0.070 &  0.079 & 0.063\\
 76 & -17.2 &  220 &   6 & 122 &  10 & -0.052 & 0.042 &  0.025 & 0.055\\
 76 & -15.5 &  224 &   6 & 112 &  10 & -0.110 & 0.044 &  0.100 & 0.063\\
 76 & -13.9 &  198 &   6 & 128 &   9 & -0.044 & 0.040 &  0.051 & 0.048\\
 76 & -12.7 &  186 &   8 & 143 &  12 & -0.061 & 0.045 &  0.110 & 0.054\\
 76 & -11.9 &  196 &   8 & 146 &  11 & -0.009 & 0.044 &  0.083 & 0.045\\
 76 & -11.0 &  160 &   7 & 161 &   8 &  0.057 & 0.036 & -0.019 & 0.039\\
 76 & -10.2 &  173 &   7 & 122 &  10 & -0.005 & 0.052 &  0.023 & 0.049\\
 76 &  -9.4 &  150 &   6 & 180 &   9 & -0.031 & 0.029 &  0.013 & 0.036\\
 76 &  -8.6 &  142 &   6 & 156 &   9 & -0.079 & 0.031 &  0.089 & 0.040\\
 76 &  -7.8 &  131 &   6 & 158 &  10 & -0.009 & 0.035 &  0.107 & 0.040\\
 76 &  -6.9 &  100 &   6 & 176 &   9 &  0.032 & 0.027 &  0.126 & 0.034\\
 76 &  -6.1 &  116 &   5 & 164 &   8 & -0.053 & 0.025 &  0.106 & 0.032\\
 76 &  -5.3 &  100 &   5 & 199 &   7 & -0.015 & 0.022 &  0.046 & 0.025\\
 76 &  -4.5 &   96 &   4 & 194 &   6 & -0.021 & 0.019 &  0.092 & 0.023\\
 76 &  -3.7 &   84 &   4 & 207 &   6 & -0.012 & 0.016 &  0.070 & 0.019\\
 76 &  -2.8 &   63 &   4 & 235 &   5 & -0.044 & 0.013 &  0.058 & 0.016\\
 76 &  -2.0 &   62 &   3 & 239 &   4 & -0.020 & 0.011 &  0.067 & 0.014\\
 76 &  -1.2 &   46 &   3 & 264 &   4 & -0.029 & 0.009 &  0.061 & 0.011\\
 76 &  -0.4 &   25 &   3 & 274 &   4 &  0.004 & 0.009 &  0.057 & 0.010\\
 76 &   0.4 &  -21 &   3 & 291 &   4 & -0.002 & 0.008 &  0.069 & 0.010\\
 76 &   1.3 &  -38 &   3 & 280 &   4 &  0.019 & 0.009 &  0.078 & 0.011\\
 76 &   2.1 &  -70 &   3 & 251 &   4 &  0.025 & 0.010 &  0.082 & 0.012\\
 76 &   2.9 &  -86 &   3 & 229 &   5 &  0.042 & 0.013 &  0.074 & 0.015\\
 76 &   3.7 & -103 &   4 & 203 &   6 &  0.044 & 0.015 &  0.107 & 0.019\\
 76 &   4.5 &  -94 &   4 & 192 &   5 & -0.003 & 0.017 &  0.074 & 0.020\\
 76 &   5.4 &  -98 &   4 & 181 &   5 & -0.055 & 0.020 & -0.011 & 0.022\\
 76 &   6.2 & -108 &   5 & 178 &   7 & -0.021 & 0.023 &  0.060 & 0.026\\
 76 &   7.0 & -110 &   5 & 156 &   7 &  0.018 & 0.026 &  0.052 & 0.030\\
 76 &   7.8 & -129 &   5 & 148 &   7 & -0.002 & 0.030 &  0.017 & 0.031\\
 76 &   8.6 & -136 &   5 & 142 &   6 & -0.030 & 0.031 & -0.013 & 0.033\\
 76 &   9.5 & -154 &   6 & 127 &   9 & -0.047 & 0.037 &  0.057 & 0.049\\
 76 &  10.3 & -166 &   6 & 158 &   9 & -0.056 & 0.033 &  0.037 & 0.040\\
 76 &  11.1 & -176 &   6 & 125 &   9 & -0.006 & 0.043 &  0.044 & 0.044\\
 76 &  11.9 & -188 &   6 & 143 &   9 &  0.034 & 0.036 &  0.024 & 0.045\\
 76 &  12.7 & -182 &   7 & 125 &  10 &  0.020 & 0.051 &  0.043 & 0.050\\
 76 &  13.6 & -206 &   7 & 141 &  12 & -0.081 & 0.043 &  0.127 & 0.054\\
 76 &  14.4 & -196 &   8 & 112 &  11 & -0.024 & 0.057 &  0.083 & 0.059\\
 76 &  15.6 & -200 &   6 & 101 &   9 &  0.023 & 0.054 &  0.090 & 0.060\\
 76 &  17.2 & -207 &   8 &  89 &   9 & -0.013 & 0.070 &  0.083 & 0.061\\
 76 &  18.9 & -209 &   6 &  73 &  14 & -0.035 & 0.059 &  0.009 & 0.115\\
 76 &  20.5 & -204 &   7 &  84 &  10 &  0.015 & 0.068 &  0.045 & 0.071\\
 76 &  22.2 & -209 &   6 &  92 &  13 & -0.063 & 0.047 &  0.013 & 0.090\\
 76 &  23.8 & -201 &   6 &  83 &  14 & -0.082 & 0.052 &  0.016 & 0.111\\
 76 &  25.4 & -214 &  10 &  78 &  10 &  0.009 & 0.100 &  0.033 & 0.069\\
 76 &  27.1 & -238 &   6 &  80 &  15 &  0.078 & 0.055 &  0.011 & 0.120\\
 76 &  28.7 & -209 &  12 &  58 &  10 & -0.004 & 0.152 & -0.033 & 0.064\\
 76 &  30.8 & -211 &   6 &  76 &  18 &  0.036 & 0.058 & -0.008 & 0.147\\
 76 &  33.2 & -215 &   6 &  63 &  24 &  0.052 & 0.060 &  0.004 & 0.213\\
 76 &  36.1 & -212 &   7 &  68 &  21 & -0.059 & 0.066 &  0.015 & 0.184\\
 76 &  39.3 & -209 &  12 &  88 &  12 & -0.000 & 0.118 &  0.038 & 0.063\\
 76 &  43.0 & -213 &  26 &  42 &  54 &  0.018 & 0.410 & -0.006 & 0.557\\
 76 &  47.8 & -183 &   8 &  58 &  44 & -0.019 & 0.075 &  0.019 & 0.389\\
 76 &  55.7 & -159 &  18 &  76 &  36 &  0.067 & 0.177 & -0.058 & 0.282\\
164 & -19.0 &   10 &  25 &  98 &  30 & -0.015 & 0.208 &  0.021 & 0.157\\
164 & -14.2 &   26 &  10 & 122 &  18 &  0.107 & 0.052 &  0.001 & 0.101\\
164 & -11.9 &   15 &  10 & 153 &  14 & -0.039 & 0.055 &  0.077 & 0.057\\
164 & -10.2 &   -7 &   7 & 121 &  10 & -0.005 & 0.052 &  0.025 & 0.054\\
164 &  -9.0 &  -18 &   7 & 132 &  12 &  0.045 & 0.045 &  0.035 & 0.058\\
164 &  -8.2 &  -12 &   6 & 138 &  10 &  0.014 & 0.040 &  0.141 & 0.049\\
164 &  -7.4 &   11 &   5 & 137 &   9 & -0.071 & 0.032 &  0.140 & 0.045\\
164 &  -6.6 &    1 &   5 & 153 &   8 & -0.024 & 0.028 &  0.088 & 0.035\\
164 &  -5.8 &    3 &   4 & 160 &   7 & -0.076 & 0.023 &  0.081 & 0.031\\
164 &  -5.0 &    5 &   4 & 175 &   6 & -0.041 & 0.020 &  0.092 & 0.026\\
164 &  -4.1 &  -12 &   3 & 197 &   5 &  0.039 & 0.015 &  0.081 & 0.018\\
164 &  -3.3 &   -8 &   3 & 225 &   5 &  0.026 & 0.012 &  0.088 & 0.015\\
164 &  -2.5 &   -2 &   3 & 250 &   4 &  0.007 & 0.010 &  0.062 & 0.012\\
164 &  -1.7 &   -2 &   3 & 262 &   4 & -0.010 & 0.009 &  0.066 & 0.011\\
164 &  -0.8 &    7 &   3 & 276 &   4 & -0.005 & 0.009 &  0.062 & 0.010\\
164 &  -0.0 &    7 &   3 & 280 &   4 & -0.003 & 0.008 &  0.082 & 0.010\\
164 &   0.8 &   12 &   3 & 278 &   4 & -0.007 & 0.008 &  0.083 & 0.010\\
164 &   1.6 &    3 &   3 & 272 &   4 & -0.011 & 0.009 &  0.097 & 0.011\\
164 &   2.4 &    4 &   3 & 254 &   4 &  0.003 & 0.010 &  0.081 & 0.012\\
164 &   3.2 &    1 &   3 & 229 &   5 & -0.010 & 0.012 &  0.078 & 0.015\\
164 &   4.1 &    6 &   3 & 202 &   5 & -0.030 & 0.015 &  0.035 & 0.017\\
164 &   4.9 &    1 &   4 & 187 &   6 & -0.018 & 0.018 &  0.079 & 0.022\\
164 &   5.7 &    4 &   4 & 170 &   7 & -0.003 & 0.022 &  0.085 & 0.028\\
164 &   6.5 &    3 &   5 & 148 &   8 &  0.025 & 0.027 &  0.111 & 0.036\\
164 &   7.4 &    3 &   6 & 146 &   9 &  0.001 & 0.034 &  0.072 & 0.039\\
164 &   8.8 &    7 &   7 & 136 &  10 & -0.002 & 0.043 &  0.051 & 0.044\\
164 &   9.0 &   -9 &   8 & 140 &  13 & -0.087 & 0.046 &  0.076 & 0.061\\
164 &  10.2 &   10 &   8 & 123 &  11 & -0.034 & 0.053 &  0.092 & 0.052\\
164 &  11.8 &   -1 &  13 & 137 &  16 & -0.018 & 0.079 &  0.050 & 0.059\\
164 &  14.1 &  -20 &  11 & 111 &  21 &  0.088 & 0.062 & -0.009 & 0.129\\
164 &  19.4 &  -48 &  20 & 140 &  32 & -0.080 & 0.068 & -0.016 & 0.172\\

\hline

\end{longtable}
\newpage
\begin{center}
\begin{table}
\caption{ionized gas kinematics along the major (PA=76\degr ) and minor 
(PA=164\degr ) axis of IC~5181.}
\begin{tabular}{rrrrrr}
\hline
\multicolumn{1}{c}{PA}       &    
\multicolumn{1}{c}{r}        &    
\multicolumn{2}{c}{V$\pm$dV} &
\multicolumn{2}{c}{$\sigma\pm$ d$\sigma$}\\

\multicolumn{1}{c}{\degr}    &    
\multicolumn{1}{c}{\arcsec}  &    
\multicolumn{2}{c}{\kms}     &
\multicolumn{2}{c}{\kms}     \\

\hline
 76 & -19.6 &  -19 &  10 &  62 &  25\\ 
 76 & -17.2 &  -14 &   8 &  84 &  25\\
 76 & -14.7 &  -21 &   6 &  73 &  14\\
 76 & -12.2 &  -45 &   7 &  69 &  18\\
 76 &  -9.8 &  -37 &  18 & 121 &  57\\
 76 &  -7.3 &   41 &  25 & 131 &  30\\
 76 &  -4.9 &  -53 &   8 & 182 &  17\\
 76 &  -3.2 &  -64 &   5 & 172 &   9\\
 76 &  -2.4 &  -29 &   2 & 167 &   4\\
 76 &  -1.6 &   -5 &   1 & 169 &   2\\
 76 &  -0.8 &    7 &   1 & 170 &   1\\
 76 &   0.1 &   15 &   1 & 178 &   2\\
 76 &   0.9 &   27 &   2 & 167 &   3\\
 76 &   1.7 &   20 &   5 & 258 &  10\\
 76 &   3.3 &  -14 &   8 & 183 &   7\\
 76 &   5.8 &   69 &  14 &  81 &  20\\
 76 &   8.3 &   12 &  55 & 112 &  44\\
 76 &  10.7 &   20 &  20 &  31 &  20\\
 76 &  18.9 & -112 &  18 &  42 &  30\\
164 & -39.0 &  182 &  32 &  46 &  46\\
164 & -33.2 &  198 &  25 &  37 &  42\\
164 & -27.5 &  185 &  30 &  64 &  51\\
164 & -21.7 &  178 &  25 &  57 &  51\\
164 & -17.6 &  196 &  25 &  64 &  28\\
164 & -15.2 &  201 &  22 &  36 &  33\\
164 & -12.7 &  197 &  15 &  66 &  23\\ 
164 & -10.3 &  136 &  18 & 121 &  23\\ 
164 &  -7.8 &  126 &  31 & 156 &  26\\ 
164 &  -6.2 &  192 &  20 & 112 &  24\\ 
164 &  -5.3 &  153 &  17 &  99 &  20\\ 
164 &  -4.5 &  169 &   7 &  87 &   8\\ 
164 &  -3.7 &  187 &   4 &  97 &   4\\ 
164 &  -2.9 &  158 &   4 & 132 &   4\\ 
164 &  -2.1 &  119 &   2 & 160 &   2\\ 
164 &  -1.2 &   79 &   1 & 171 &   2\\ 
164 &  -0.4 &   22 &   1 & 182 &   1\\ 
164 &   0.4 &  -35 &   1 & 173 &   1\\ 
164 &   1.2 &  -97 &   2 & 186 &   4\\ 
164 &   2.0 & -132 &   2 & 166 &   5\\ 
164 &   2.9 & -161 &   2 & 138 &   4\\ 
164 &   3.7 & -152 &   5 & 114 &   7\\ 
164 &   5.3 & -142 &   7 & 108 &   9\\ 
164 &   7.8 & -155 &  12 &  75 &  17\\ 
164 &  10.2 & -176 &  13 &  66 &  24\\ 
164 &  12.7 & -192 &  14 &  55 &  22\\ 
164 &  15.2 & -203 &  13 &  62 &  29\\ 
164 &  18.4 & -137 &  32 &  68 &  48\\ 
164 &  25.0 &  -53 & 108 & 141 &  79\\ 
164 &  34.0 & -126 &  52 &  53 &  51\\
\hline                              
\end{tabular}
\end{table}
\end{center}

\newpage
\begin{center}
\begin{table}
\caption{Line strength measurements along the major (PA=76\degr ) and minor 
(PA=164\degr ) axis of IC~5181.}
\begin{tabular}{rrrrrrrrrrrr}
\hline 
\multicolumn{1}{c}{PA}   &    
\multicolumn{1}{c}{r}    &    
\multicolumn{2}{c}{\Hb}  &
\multicolumn{2}{c}{\Mg2} &
\multicolumn{2}{c}{\Mgb} &
\multicolumn{2}{c}{{\rm Fe}$_{5270}$}  &
\multicolumn{2}{c}{{\rm Fe}$_{5335}$}\\

\multicolumn{1}{c}{\degr}        &    
\multicolumn{1}{c}{\arcsec}      &    
\multicolumn{2}{c}{\AA$\pm$d\AA} &
\multicolumn{2}{c}{mag$\pm$dmag} &
\multicolumn{2}{c}{\AA$\pm$d\AA} &
\multicolumn{2}{c}{\AA$\pm$d\AA} &
\multicolumn{2}{c}{\AA$\pm$d\AA}\\
\hline
 76 & -40.6 &  1.76 &  0.18 & 0.243 & 0.005 &  3.86 &  0.17 &  2.41 &    0.19 &   3.23 &    0.20\\
 76 & -26.3 &  2.16 &  0.14 & 0.253 & 0.003 &  4.30 &  0.13 &  2.27 &    0.14 &   2.67 &    0.15\\
 76 & -18.7 &  1.94 &  0.12 & 0.262 & 0.003 &  4.62 &  0.12 &  2.57 &    0.13 &   2.72 &    0.14\\
 76 & -13.0 &  1.59 &  0.12 & 0.266 & 0.003 &  4.63 &  0.11 &  2.93 &    0.12 &   2.56 &    0.13\\
 76 &  -8.9 &  1.78 &  0.11 & 0.259 & 0.003 &  4.43 &  0.11 &  2.71 &    0.12 &   2.71 &    0.13\\
 76 &  -6.0 &  1.66 &  0.09 & 0.278 & 0.003 &  4.97 &  0.10 &  3.17 &    0.11 &   2.99 &    0.12\\
 76 &  -4.0 &  1.84 &  0.09 & 0.291 & 0.003 &  5.01 &  0.09 &  3.09 &    0.10 &   3.12 &    0.12\\
 76 &  -2.8 &  1.72 &  0.10 & 0.302 & 0.003 &  5.50 &  0.11 &  3.36 &    0.12 &   3.14 &    0.14\\
 76 &  -2.0 &  1.35 &  0.10 & 0.305 & 0.003 &  5.16 &  0.11 &  3.26 &    0.12 &   3.11 &    0.14\\
 76 &  -1.2 &  1.11 &  0.10 & 0.320 & 0.003 &  5.83 &  0.10 &  3.34 &    0.11 &   3.82 &    0.14\\
 76 &  -0.4 &  1.56 &  0.10 & 0.334 & 0.003 &  5.91 &  0.11 &  3.37 &    0.12 &   3.80 &    0.15\\
 76 &   0.4 &  0.97 &  0.11 & 0.338 & 0.003 &  5.97 &  0.12 &  3.43 &    0.12 &   3.62 &    0.16\\
 76 &   1.3 &  1.18 &  0.11 & 0.329 & 0.003 &  6.02 &  0.12 &  3.27 &    0.12 &   3.60 &    0.15\\
 76 &   2.1 &  1.48 &  0.11 & 0.321 & 0.003 &  5.84 &  0.11 &  3.43 &    0.12 &   3.18 &    0.14\\
 76 &   3.0 &  1.81 &  0.11 & 0.315 & 0.003 &  5.58 &  0.12 &  3.39 &    0.12 &   3.25 &    0.14\\
 76 &   3.7 &  1.50 &  0.11 & 0.298 & 0.003 &  5.16 &  0.12 &  3.18 &    0.12 &   3.24 &    0.14\\
 76 &   5.0 &  1.66 &  0.10 & 0.284 & 0.003 &  4.94 &  0.10 &  3.12 &    0.10 &   3.20 &    0.12\\
 76 &   6.9 &  1.88 &  0.10 & 0.269 & 0.003 &  4.94 &  0.10 &  2.90 &    0.10 &   2.87 &    0.12\\
 76 &   9.8 &  1.58 &  0.10 & 0.260 & 0.003 &  4.54 &  0.11 &  2.78 &    0.11 &   3.01 &    0.13\\
 76 &  13.8 &  2.25 &  0.10 & 0.265 & 0.004 &  4.56 &  0.12 &  2.96 &    0.11 &   2.83 &    0.13\\
 76 &  19.6 &  2.03 &  0.10 & 0.263 & 0.004 &  4.33 &  0.12 &  2.97 &    0.11 &   2.65 &    0.13\\
 76 &  27.2 &  1.88 &  0.11 & 0.253 & 0.004 &  4.04 &  0.13 &  2.78 &    0.12 &   2.37 &    0.14\\
 76 &  41.5 &  1.61 &  0.13 & 0.216 & 0.005 &  3.56 &  0.17 &  2.58 &    0.15 &   2.45 &    0.19\\
164 & -11.9 &  2.09 &  0.16 & 0.242 & 0.007 &  4.86 &  0.23 &  2.28 &    0.30 &   2.64 &    0.28\\
164 &  -6.2 &  1.33 &  0.07 & 0.269 & 0.003 &  4.67 &  0.09 &  2.99 &    0.11 &   2.72 &    0.11\\
164 &  -4.0 &  1.63 &  0.06 & 0.286 & 0.003 &  5.02 &  0.09 &  2.99 &    0.11 &   2.63 &    0.11\\
164 &  -2.8 &  1.45 &  0.06 & 0.308 & 0.003 &  5.28 &  0.10 &  3.11 &    0.12 &   3.24 &    0.12\\
164 &  -2.0 &  1.13 &  0.06 & 0.312 & 0.003 &  5.38 &  0.10 &  3.29 &    0.12 &   3.07 &    0.12\\
164 &  -1.2 &  1.09 &  0.07 & 0.327 & 0.003 &  5.91 &  0.10 &  3.23 &    0.12 &   3.30 &    0.13\\
164 &  -0.3 &  0.95 &  0.07 & 0.335 & 0.003 &  5.93 &  0.11 &  3.33 &    0.13 &   3.56 &    0.14\\
164 &   0.5 &  1.00 &  0.08 & 0.334 & 0.003 &  5.86 &  0.11 &  3.55 &    0.14 &   3.49 &    0.14\\
164 &   1.4 &  1.05 &  0.08 & 0.328 & 0.003 &  5.91 &  0.11 &  3.42 &    0.13 &   3.51 &    0.14\\
164 &   2.1 &  1.05 &  0.08 & 0.322 & 0.003 &  5.74 &  0.11 &  3.62 &    0.13 &   3.30 &    0.14\\
164 &   2.9 &  1.07 &  0.08 & 0.307 & 0.003 &  5.57 &  0.11 &  3.33 &    0.13 &   3.04 &    0.13\\
164 &   3.8 &  1.35 &  0.09 & 0.300 & 0.003 &  5.33 &  0.12 &  3.19 &    0.14 &   3.46 &    0.14\\
164 &   4.9 &  1.94 &  0.08 & 0.283 & 0.003 &  4.91 &  0.11 &  2.97 &    0.13 &   3.08 &    0.12\\
164 &   6.9 &  1.83 &  0.08 & 0.265 & 0.003 &  4.64 &  0.11 &  3.26 &    0.13 &   2.58 &    0.12\\
164 &  13.5 &  4.39 &  0.23 & 0.216 & 0.008 &  5.11 &  0.28 &  2.67 &    0.34 &   1.39 &    0.32\\
\hline                                                                                                
\end{tabular}
\end{table}                                                                                   
\end{center}
                                                                                 
\end{document}